\documentclass[aps,pre,twocolumn,superscriptaddress,showpacs]{revtex4-1}
\usepackage{amsmath}
\usepackage{amssymb}
\usepackage{graphicx}% Include figure files
\usepackage{dcolumn}% Align table columns on decimal point
\usepackage{bm}% bold math

\newcommand{\refsec}[1]{\mbox{Sec.~\ref{#1}}}
\newcommand{\reftab}[1]{\mbox{Tab.~\ref{#1}}}
\def\be{\begin{equation}}
\def\ee{\end{equation}}
\def\bea{\begin{eqnarray}}
\def\eea{\end{eqnarray}}
\def\II{\hbox{$1\hskip -1.2pt\vrule depth 0pt height 1.6ex width 0.7pt\vrule depth 0pt height 0.3pt width 0.12em$}}
\renewcommand{\Re}{\mathrm{Re}}

\begin{document}\preprint{CERN-PH-TH/2015-XXX}

\title{Cross-Section Fluctuations in Chaotic Scattering Systems}
\author{Torleif E. O.   Ericson}
\email{torleif.ericson@cern.ch}
\affiliation{Theory Department, CERN, CH-1211 Geneva, Switzerland}
\author {Barbara Dietz}
\email{dietz@ikp.tu-darmstadt.de}
\affiliation{Institut f\"{u}r Kernphysik, Technische Universit\"at Darmstadt, D-64289 Darmstadt, Germany}
\affiliation{Institute of Physics, Polish Academy of Sciences, Al. Lotnik\'ow 32/46, 02-668 Warzawa, Poland}
\author {Achim Richter}
\affiliation{Institut f\"{u}r Kernphysik, Technische Universit\"at Darmstadt, D-64289 Darmstadt, Germany}
\date{\today}

\begin{abstract}
Exact analytical expressions for the cross-section correlation functions of chaotic scattering systems have hitherto been derived only under special conditions. The objective of the present article is to provide expressions that are applicable beyond these restrictions. The derivation is based on a statistical model of Breit-Wigner type for chaotic scattering amplitudes which has been shown to describe the exact analytical results for the scattering ($S$)-matrix correlation functions accurately. Our results are given in the energy and in the time representations and apply in the whole range from isolated to overlapping resonances. The $S$-matrix contributions to the cross-section correlations are obtained in terms of explicit irreducible and reducible correlation functions. Consequently, the model can be used for a detailed exploration of the key features of the cross-section correlations and the underlying physical mechanisms. In the region of isolated resonances, the cross-section correlations contain a dominant contribution from the self-correlation term. For narrow states the self-correlations originate predominantly from widely spaced states with exceptionally large partial width. In the asymptotic region of well-overlapping resonances, the cross-section autocorrelation functions are given in terms of the $S$-matrix autocorrelation functions. For inelastic correlations, in particular, the Ericson fluctuations rapidly dominate in that region. Agreement with known analytical and with experimental results is excellent.
\end{abstract}
\maketitle
\section {Introduction \label{Sec:Introduction}}
Scattering processes from highly complex systems show characteristic fluctuation phenomena~\cite{Guhr1998,Mitchell2009,MIT10}.  Their average features  are of generic nature independently of the details of the underlying interactions and show sensitivity to the energy or frequency. Such fluctuations are characteristic for quantum chaotic scattering, 
 which occurs in nuclei~\cite{ERI66,BRO81}, molecules, in the conductance of mesoscopic semiconductor devices~\cite{Jalabert1990} and disordered open systems~\cite{BEE08}, in electronic transport in ballistic open quantum dots~\cite{BEE97,Alhassid2000} and in microwave cavities~\cite{Doron1990,Schaefer2003,DIE08}. Generally speaking a scattering process is chaotic, if the scattered particle is trapped, i.e., moves close to the periodic orbits of the corresponding closed system for a sufficiently long time in the interaction region so that it is effected by the interior dynamics, which is required to be chaotic~\cite{Bluemel1992,Jung1997}. A lower bound for this time, and therewith an upper bound for the energy spacings may be obtained based on a semiclassical approach and is given by the period of the shortest periodic orbit in the corresponding closed classical system~\cite{Berry1985}. For a semiclassical treatment of nuclei see, e.g., Refs.~\cite{Brack1972,Strutinsky1976,Bohigas2002}. In molecules chaotic scattering has been studied in indirect photodissociations of excited molecules through resonances, i.e., quasibound states formed due to the presence of a potential barrier which hinders the immediate dissociation~\cite{Alhassid1998}. Similarly, the statistical theory of compound nucleus reactions relies on the formation of a compound nucleus as an intermediate state. Analytical results were obtained for the fluctuation properties of the corresponding scattering matrix~\cite{VER85} based on the ansatz of Bohr~\cite{Bohr1936}, that the formation and decay of the compound nucleus are independent processes. The associated cross sections~\cite{MIT10,ERI66,BRO81,Schaefer2003,DIE08,DIE10A} exhibit random fluctuations as a function of energy, thereby reflecting the randomness in the individual components of its resonant structure. The ratio $\Gamma _W /d$ of the average total resonance width $\Gamma _W $ and the average resonance spacing $d$  characterizes the energy (frequency) regions ranging from isolated resonances with $\Gamma _W\ll d$ to the overlapping ones with $\Gamma _W\gtrsim d$. Data give information on the average cross section, their variance and the correlation functions, which are the major tools for the investigation of the resonance structure of scattering systems.  \\
A general theoretical treatment of chaotic scattering processes has been developed by Verbaarschot-Weidenm\"{u}ller-Zirnbauer (VWZ) combining scattering theory and random matrix theory (RMT)~\cite{VER85,PLU12}. The results are expressed in terms of the values of $d$ and the average elements $\langle S\rangle$ of the scattering matrix $S$. These are the characteristic physical parameters. The theory gives global analytical results for the autocorrelation and cross-correlation functions of an $S$-matrix element and a complex conjugate, one as well as for its third and fourth moment~\cite{DAV88,DAV89}. Reference~\cite{DAV88}, in particular, provides analytical expressions for the average cross-section coefficient~\cite{SAT63} and its variance. The accuracy of these analytical results was tested thoroughly with RMT simulations and also in experiments with microwave billiards~\cite{DIE10,DIE10A}. In general, however, cross-section autocorrelations have a more complex structure and are not known analytically. Explicit results could so far only be obtained using numerical simulations based on RMT~\cite{DIE10}.

The present article aims at filling this gap by providing an analytic, albeit an approximate,  description of the cross-section correlations based on a $S$-matrix model, using the traditional statistical Breit-Wigner (SBW) resonance model inspired by nuclear reaction theory~\cite{ERI60,ERI63,ERI66,GOR02}.  It is also referred to as the EGS model in Ref.~\cite{ERI13} or the rescaled Breit-Wigner model in Ref.~\cite{GOR02}. The SBW model approximates the $S$ matrix by a coherent sum of resonances with random partial width amplitudes of appropriate average strength and a total width given by the sum of the partial widths associated with the open channels. It yields a remarkably good description for the $S$-matrix autocorrelations~\cite{ERI13}. This feature of the SBW model was the motivation for its application to the cross-section correlations in the present article. Approximations and predictions~\cite{AGA75,VER86,HAR87,HAR90,MUE90,DIT00,Fyodorov2005,DIE10B} for the latter including an extensive comparison to experimental data exist for the asymptotic regions of isolated and strongly overlapping resonances~\cite{ERI66,Lyn68,ERI60,Eri63,Blu88,ALT95,Schaefer2003,DIE08,Lawniczak2008}. We first extend them by deriving the SBW expressions for the $S$-matrix and cross-section autocorrelation functions both in the energy and in the time representation. Their usefulness and accuracy is then established numerically by comparison with known exact results derived within RMT on the basis of the supersymmetry method~\cite{DAV88,DAV89,DIE10}. The very good agreement of these results with those derived with the SBW model gives confidence that it describes the physics correctly. This is corroborated with RMT simulations. In~\refsec{Exp}, our results are checked with experimental data obtained from measurements with microwave billiards. Finally, the SBW model provides a detailed insight into the origin of the contributions of various $S$-matrix correlation functions, which are not accessible within the VWZ model. It allows us to investigate the transition region from isolated resonances to overlapping ones separately for the irreducible two-, three- and four-point correlations contributing to the cross-section correlations. Furthermore, we demonstrate explicitly the central r\^ole played by the channels for which the scattering signal is recorded.  
\section{Framework\label{Sec:Framework} }
\subsection {From Cross-Sections to $S$-Matrix Correlations\label{Sec:Decomposition}}
We consider chaotic scattering in a time-reversal invariant system described by a unitary and symmetric scattering matrix $S_{ab}(E)$. Here $a,b=1,\cdots ,\Lambda$ denotes the channels and $E$ the energy. The classical dynamics of the closed system is assumed to be chaotic. Accordingly, its spectral properties are described by random matrices of large dimension from the Gaussian orthogonal ensemble (GOE). The associated scattering system exhibits a large number of resonances that are coupled dynamically to the channels as described for example in Ref.~\cite{DIE10A}. We are mainly interested in the fluctuation properties of the $S$-matrix elements and the cross-sections $\sigma _{ab}(E)=|S_{ab}(E)-\delta _{ab}|^2$, which are analyzed in terms of correlation functions. These are obtained as ensemble averages generated by random variables, which are equivalent to energy averages in the absence of secular variations. For definitions and notations we follow with some minor changes the procedure and notations of Ref.~\cite{DIE10A}, with a brief reminder below. As in Ref.~\cite{DIE10} we limit the discussion to the cross-section autocorrelations for simplicity of presentation. The SBW results are readily generalized to other cases.

The basic quantity is the scattering matrix $S(E)$. It is convenient to decompose it into an average and a fluctuating part~\cite{ENG73,WEI84}, 
\begin{equation}
S_{ab}(E)=\langle S_{aa}\rangle \delta _{ab}+S_{ab}^{fl}(E)\label{saverage},
\end{equation}
with $\langle S^{fl}_{ab}\rangle =0$. Averages are indicated by brackets as $\langle\cdots\rangle $. The $S$-matrix autocorrelation function is defined as 
\begin{eqnarray}
\label{Eq:definitions}
C^{(2)}_{ab}(\epsilon)&&=\langle S_{ab}^{fl}(E-\epsilon /2)S_{ab}^{fl*}(E+\epsilon/2)\rangle\\
&&\equiv \langle S_{ab}(E-\epsilon /2)S_{ab}^*(E+\epsilon/2)\rangle - \left |\langle S_{ab}(E) \rangle \right |^2,
\nonumber
\end{eqnarray} 
where we introduced the abbreviation $C^{(2)}_{ab}(\epsilon)=C[S_{ab}^{*} S_{ab}^{}](\epsilon )$.
The channels corresponding to the labels $a,b$ in the $S$-matrix autocorrelation function are denoted as the observed channels in the sequel. The parameters of $S(E)$ are the average resonance spacing $d$ and the transmission coefficients in all open channels $e$,
\begin{equation}
T_e=1-|\langle S_{ee}\rangle |^2.\label{transm}
\end{equation}
These quantities determine the scale of the average correlation width $\Gamma _W $ of the $S$-matrix in terms of the Weisskopf estimate~\cite{Blatt1952}
\begin{equation}
\Gamma _W=\sum_e\langle\Gamma_e\rangle =\frac{d}{2\pi}\sum_eT_e.\label{weiss}
\end{equation}
The average cross section is given by $\langle \sigma _{ab}(E)\rangle =\langle |S_{ab}(E)-\delta_{ab}|^2 \rangle =|\langle S_{ab}\rangle-\delta_{ab}|^2+\langle |S_{ab}^{fl}|^2\rangle$. The fluctuations of the cross sections are described by their autocorrelation functions. Here we should note that Ref.~\cite{DIE10A} defines the cross section as $\sigma_{ab}=|S_{ab}|^2$. It differs from our definition $\sigma_{ab}=|S_{ab}-\delta_{ab}|^2$ which is the one commonly used in nuclear physics. The cross-section autocorrelation function is then obtained as
\begin{eqnarray}
\label{Eq:defcrosscorrfunc}
C_{ab}(\epsilon) =&&\langle |S_{ab}(E+\epsilon /2)-\delta_{ab}|^2 |S_{ab}(E-\epsilon /2)
-\delta_{ab}|^2 \rangle\nonumber \\
&&- \langle |S_{ab}-\delta_{ab}|^2 \rangle^2.
\end{eqnarray}
Note, that we use the abbreviation $C_{ab}(\epsilon)=C[\sigma _{ab}\sigma _{ab}](\epsilon)$ for the cross-section autocorrelation functions. 
This function is the primary object of the studies in the present article. It can be decomposed as follows
 \begin{eqnarray}
  &&C_{ab}(\epsilon) \label{Eq:4pointcorrgeneralcrossA} =\\
  &&2\delta_{ab}\Re\Big\{(1-\langle S_{aa} \rangle)^2C^{(2)}_{aa}(\epsilon)\label{Eq:2pointSA}\\
  +&&(1-\langle S_{aa}\rangle) \left\langle\left[ S^{fl*}_{aa}(E-\epsilon )\nonumber
  +S^{fl*}_{aa}(E+\epsilon)\right]|S_{aa}^{fl}(E)|^2~\right\rangle\Big\}\\
  \label{Eq:fluctuation}
  +&&|C^{(2)}_{ab}(\epsilon  )|^2\\
  +&&\Big\{\left \langle |S_{ab}^{fl} (E-\epsilon /2 )|^2 |S_{ab}^{fl}(E+\epsilon /2 )|^2 \right \rangle\label{Eq:4pointfl} - \left \langle |S_{ab}^{fl}|^2 \right \rangle ^2\\
-&& \left\vert\left\langle  S_{ab}^{fl} (E-\epsilon /2)S_{ab}^{fl*} (E+\epsilon /2 )\right\rangle\right\vert^2\Big\}.\nonumber  
\end{eqnarray}
Here, the average $S$-matrix is assumed to be diagonal and real~\cite{VER85}, $\langle S_{a b} \rangle =0$ for $a\ne b$. In the decomposition above the four-point term~(\ref{Eq:4pointfl}) is of particular importance. It is denoted by
\begin{eqnarray}
&&\label{Eq:C^4(epsilon)} C_{ab}^{(4)}(\epsilon)=\\
&&\left \langle |S_{ab}^{fl} (E-\epsilon /2 )|^2 |S_{ab}^{fl}(E+\epsilon /2 |^2 \right \rangle  
-\left \langle |S_{ab}^{fl}|^2 \right \rangle^2.\nonumber 
\end{eqnarray}
Further decomposition of this expression yields terms of the type $\langle  S_{ab}^{fl}(E_1)S_{ab}^{fl} (E_2) \rangle $ and its conjugate. They vanish on performing the energy average, since the poles of the two matrix elements are in the same half of the complex energy plane. The only surviving product of such pairs in~(\ref{Eq:C^4(epsilon)}) is the $S$-matrix correlator~(\ref{Eq:definitions}) multiplied by its conjugate, $|C^{(2)}_{ab}(\epsilon)|^2 $. For ineleastic scattering $(a\ne b)$ this term is identical to the square of the $S$-matrix correlation function -- the Ericson fluctuation term -- which  dominates   the region of overlapping resonances. In view of its importance it has been explicitly subtracted in~(\ref {Eq:4pointfl}) and added as~(\ref{Eq:fluctuation}). Therefore, the term in curly brackets in~(\ref {Eq:4pointfl}) depends only on the averages of products of four $S$-matrix elements and cannot be decomposed into simpler averages. This is usually referred to as an irreducible part of the correlations. It is dominant for isolated resonances and provides an important contribution in the transition region towards overlapping levels. Its properties are a major topic of the present article.
 
 The elastic three-point term in the curly brackets of~(\ref{Eq:2pointSA}) is denoted by 
\begin{equation}
 C_{ab}^{(3)} (\epsilon)= \langle  S ^{fl*}_{aa}(E+\epsilon)   |S_{aa}^{fl}(E)|^2 \rangle\delta_{ab}. \label{Eq:3pointflA} 
\end{equation}
The decomposition of the cross-section correlations  $C_{ab}(\epsilon )$ above  is general.  
\subsection{The Statistical Breit-Wigner Model \label{Sec:SBW} }
We have chosen a  resonance model which has been extensively used in the past with minor variations for investigations of statistical reaction properties, mainly in the asymptotic regions of isolated or strongly overlapping levels~\cite{ERI60,ERI63,ERI66,GOR02,GOR14}. It provides an approximate model for the description of the properties of the $S$-matrix correlations contributing to the cross-section correlation functions; see  Eqs.~(\ref{Eq:4pointcorrgeneralcrossA})-(\ref{Eq:4pointfl}). For convenience we refer to this model as the SBW model. It has its roots in nuclear scattering theory, e.g., in Feshbach's unified model~\cite{FES62}, and is formulated so as to be valid both for isolated and for overlapping resonances~\cite{ERI60,ERI63,ERI66,GOR02}. We assume a situation of scattering from a complex, initially closed system which hosts a large number of states, that are coupled to a set of uncorrelated open channels. This produces poles at complex energies $e_k$ with  Breit-Wigner type pole contributions to the resulting $S$-matrix. The coupling to the open channels $e$ is described by statistically distributed partial width amplitudes $\gamma _{e k}$ characteristic of each pole $k$. The sum over the partial widths $\Gamma_{e k}=\gamma_{e k}^2$ yields the total width $\Gamma _k=\sum_{e }\Gamma _{e k}$. The average resonance spacing $d$ as well as the partial width amplitudes $\gamma _{e k}$ are assumed to have no long-range secular variation with energy and the channel thresholds are required to be far larger than $d$. The states form a statistical ensemble with average properties. 

More precisely, we define the resonant contribution $ S_{ab}^{res}(E)$  to the scattering matrix  as a sum of uncorrelated   resonances, which are unitary when taken individually~\cite{footnote},
\begin{eqnarray}
\label{Eq:S-abres}
S_{ab}^{res}=&&-i\sum _k{\gamma _{ak} \gamma _{kb} \over E-e_k}~,~~e_k=E_k -i\Gamma _k/2~,\\ 
\Gamma _{e k}=&&\gamma _{e k}^2~,~~ \Gamma _{ k}=\sum _{e}\Gamma _{e k}.\nonumber
\end{eqnarray}  
Here, $E_k$ denotes the position of the $k$th resonance. The averages of these quantities are obtained in terms of the transmission coefficients $T_e$ defined in Eq.~(\ref{transm}),
\begin{equation}
\label{Eq:Definitions}
\langle\Gamma _{ek}\rangle=\langle \gamma_{ek}^2 \rangle =(d/2\pi )T_e~,~~\langle\Gamma _k\rangle =(d/2\pi)\sum_e T_e.
\end{equation}  
Elastic phase shifts have been omitted as in Ref.~\cite{VER85}. The average $\langle S_{ab}^{res} \rangle $ vanishes in the non-diagonal case $(a\neq b)$ as it does for the corresponding full $S$ matrix; see Eq.~(\ref{saverage}). The average of the diagonal part is irrelevant in the present context, since it does not contribute to averages involving only $S^{fl}$. The spectral properties of the energy levels are assumed to coincide with those of random matrices from the GOE.

The central assumption is that the partial width amplitudes $\gamma _{e k}$ are random with random sign and they are commonly assumed to have a Gaussian probability distribution, such that the partial widths $\Gamma_{ek}=\gamma_{ek}^2$ have a Porter-Thomas distribution~\cite{Graf1992}; see Appendix~\ref{App:shorthand}. The motivation for these prerequisites can be expressed in several ways. One can for example view the amplitudes $\gamma _{e k}$ as the projection of the partial width operator $\hat{\gamma } _{e }$ onto a randomly oriented space spanned by a large number of orthogonal resonant states $k$ so that the sign of each partial width amplitude is random and consequently averages involving odd powers $\gamma _{e k}^{2n+1}$  vanish. This implies according to a classical statistical argument ('The Drunken Sailor Problem') that each $ \gamma _{e k}$ has a Gaussian distribution centered at the origin and is uncorrelated with the other partial width amplitudes, $ \langle \gamma _{ek}\gamma _{e^{\prime }l} \rangle  =0$ for $e\neq e^{\prime }$. The distribution of the corresponding partial widths is the Porter-Thomas distribution~\cite {POR56}. 

While a Gaussian probability distribution for the partial width amplitudes is a natural consequence of the statistical picture above, many results obtained on the basis of the model Eq.~(\ref{Eq:S-abres}) appear to be well approximated by weaker conditions assuming a symmetric distribution with random sign of the variables. Such a situation may occur for systems that are not fully chaotic. Therefore, it is of considerable interest to understand how rapidly the above established statistical limit becomes important in practice. Since the formal steps in the derivation of the cross-section autocorrelation function are identical for the case of a Porter-Thomas distribution and the general one, we consider the latter in the following and then confine ourselves to the former, when comparing to known analytical results and experimental data.

The level correlations are introduced phenomenologically and are taken to be robust. They are assumed to coincide with those of random matrices from the GOE~\cite{BRO81,MEH67}. The results, in fact, are sensitive only to their gross features: anticorrelations of levels at close encounter create a correlation hole on the typical scale $d$ corresponding to the absence of one level in that region~\cite{ALT97}. Furthermore, the energy levels are taken to be statistically independent of the partial widths. The remaining parameters of the model are the average resonance spacing and the transmission coefficients of the channels.

Stated in this form, the SBW model is completely defined and can be solved in closed form for the correlations between two, three and four {\it S}-matrix elements. 
A frequent critique concerning the SBW model is that it neglects  unitarity constraints. For a dynamical model this is justified, since consistency is essential. For the phenomenological SBW model the average $S$-matrix elements $\langle S_{aa}(E)\rangle$ depend importantly on unitarity due to the shadow of the inelastic states via the optical theorem. In particular, the inclusion of this contraint ensures that complete transmission $T_a=1$ indeed corresponds to $\langle S_{aa}(E)\rangle=0$. Other restrictions related to unitarity for the SBW model are expected to be small. The obvious quantitative success in describing many properties of chaotic systems is a strong justification for this procedure. We further remind the reader that regions of large probability are the ones least exposed to such constraints. In addition, the cross-section  correlations studied in the following result from folding procedures deemphasizing the effect of any unitarity violation in regions of low probability. 
\subsection  {$S$-Matrix Autocorrelation Function\label{Sec:2point}}
This section mainly serves to specify the notation used in the following and to  illustrate schematically the method used in~\refsec{Sec:crosscorr} for the derivation of the cross-section correlation functions within the SBW model. More complete results for the properties of its $S$-matrix autocorrelation functions are given in Appendix \ref{App:Results}.\\

The $S$-matrix autocorrelation function is defined as in the VWZ model~\cite{VER85} in order to facilitate the comparison of both models. Since $ \langle\gamma _{ki}\rangle =0$, for inelastic processes $S_{a \neq b}^{(fl ) } \equiv  S^{res}_{ab} $. Note, that correlations are non-vanishing only when averages are taken between an $S$-matrix element and a complex conjugate one. 
\subsubsection{Inelastic Autocorrelations}
We illustrate the procedure of the analysis for the inelastic autocorrelation function $C^{(2)}_{a\neq b}(\epsilon)$, which displays most features of the general case. Since $ \langle S_{ab}^{res} \rangle =0 $ for $a\ne b$,  according to Eq.~(\ref{Eq:definitions})
\begin{eqnarray}
C^{(2)}_{a\neq b}(\epsilon)=&&\left \langle S^{res}_{ab}(E-\epsilon/ 2)S^{res*}_{ab}(E+\epsilon/ 2) \right \rangle\label{Eq: C_{ab}}\\
=&& \left \langle   \sum _{kl}{\gamma _{ak} \gamma _{kb} \over (E-\epsilon/ 2-e_k)}{\gamma _{al} \gamma _{lb} \over (E+\epsilon /2-e_l^*)} \right \rangle\, . \nonumber
\end{eqnarray} 
Since the signs of the partial width amplitudes are random, only the diagonal terms with $k=l$ contribute to the average. We rescale the energy levels $E_k$ to mean spacing unity, $d=1$. Their sum is replaced by an integral over the variable $E_1$ and the label $k$ of the partial widths $\Gamma_{ek}$ is replaced by the index $1$ assuming a probability distribution $p(x_e)$ for them, which is not further specified at this point. Then Eq.~(\ref{Eq: C_{ab}}) takes the form 
\begin{eqnarray}
C^{(2)}_{a\neq b}(\epsilon)\nonumber
&&=\int_{-\infty }^{\infty}{\rm d}E_1\left\langle\frac{\Gamma _{1a}\Gamma_{1b}}{(E_1-\epsilon/ 2-e_1)(E_1+\epsilon /2-e_1^\star)}\right\rangle\\
\label{Eq:2pointcorababenergy}
&&=2\pi\left \langle {\Gamma _{a}\Gamma _{b}\over i\epsilon +\Gamma }\right\rangle\\
e_1&&=E_1-i\Gamma_1/2\nonumber,\\
\langle\cdots\rangle&&=\prod_e\int {\rm d}x_ep(x_e)\cdots .\nonumber
\end{eqnarray}
The autocorrelation function $ C^{(2)}_{a\neq b}(\epsilon =0)=\langle \sigma _{ab} \rangle $ yields the average inelastic cross section. Replacing in Eq.~(\ref{Eq:2pointcorababenergy}) each partial width $\Gamma _{e}$ by its average $\langle \Gamma _e \rangle$ reproduces the Hauser-Feshbach expression~\cite{HAU52}, also obtained within the VWZ model~\cite{MUE87,GOR98}.
 
 While the energy representation is natural in the sense that experiments are performed in it, and useful for the qualitative understanding of gross features, the time representation is far more convenient for detailed theoretical predictions. It is obtained via the Fourier transform from energy to time which vanishes for $\tau <0$. Therefore, we restrict to $\tau\geq 0$ in the following. This yields for the Fourier transform of the $S$-matrix autocorrelation function~(\ref{Eq:2pointcorababenergy})
\begin{eqnarray}
\tilde{C}^{(2)}_{a\neq b}(\tau)=&&\int_{-\infty }^{\infty }{\rm d}\epsilon \exp (2\pi i\epsilon\tau )C^{(2)}_{a\neq b}(\epsilon)\label{Eq:2-pointFouriergen}\\
=&&(2\pi )^2  \langle \Gamma_{a}\Gamma_{b  }\exp \left(-2\pi \Gamma \tau   \right) \rangle.\nonumber
\end{eqnarray}
Note, that energy is expressed in units of the level spacing $d$ (which is set to unity) and the time in units $2\pi /d$. The Fourier transform~(\ref{Eq:2-pointFouriergen}) has the great advantage, that it is separable. Thus, since the total width $\Gamma$ of a resonance is the sum of the partial widths $\Gamma_e$, the average occurring in Eq.~(\ref{Eq:2-pointFouriergen}) can be replaced by a product over averages of the individual partial widths,
\begin{eqnarray}
\tilde{C}^{(2)}_{a\neq b}(\tau)\label{Eq:2-pointFourierGamma}=&&\nonumber
(2\pi )^2  \langle \Gamma_a\exp\left(-2\pi\Gamma_a\tau\right)\rangle\langle\Gamma_b\exp\left(-2\pi\Gamma _b\tau\right)\rangle\times\nonumber\\
&&\prod_{e\neq a,b}\langle\exp\left(-2\pi \Gamma _e\tau \right)\rangle . 
\end{eqnarray}
It is particularly convenient to use a short-hand notation for the  products appearing in the Fourier transform as defined in detail in Appendix~\ref{App:shorthand}.  For the present case, Eq.~(\ref{Eq:Pi_eabc}) gives with $k_{a}=k_{b}=1$, $k_{e\neq a,b}=0$ and $x_e=\Gamma _e/\langle \Gamma _e\rangle $
\begin{eqnarray}
\tilde{C}^{(2)}_{a\neq b}(\tau )\label{Eq:2-pointFourier}=&&T_aT_b \Pi_{e; ab}(\tau )\\
\equiv&& T_aT_b \prod _{e}  \left \langle x_e^{k_{e} } \exp(-T_{e }\tau   x_e) \right \rangle .\nonumber
\end{eqnarray}
Here, as commonly done in the time-representation, the average partial widths are replaced by the transmission coefficients; see Eq.~(\ref{Eq:Definitions}). 
In the standard case of a Porter-Thomas distribution the inelastic autocorrelation function becomes according to Eq.~(\ref{Eq:shorthandsumgeneral})
\begin{eqnarray}
\label{Eq:PieabPorterThomas}
\tilde{C}^{(2)}_{a\neq b}(\tau )=&&T_aT_b \Pi_{e; ab}(\tau )\\
\equiv&& T_aT_b(1+2\tau T_a)^{-1}(1+2\tau T_b)^{-1}\prod _e(1+2\tau T_e)^{-1/2} .\nonumber 
\end{eqnarray}
 The corresponding expression for a generalized Porter-Thomas distribution is also given in Appendix~\ref{App:shorthand}. 
  
The average inelastic cross section $\langle\sigma_{ab}\rangle =C^{(2)}_{a\neq b}(0)\equiv\langle |S_{ab}^{res}|^2\rangle$ is obtained from the inverse Fourier transform of the autocorrelation function in the time-representation~(\ref {Eq:2-pointFourier}),
\begin{eqnarray}
\label {Eq:sumrule}  
\langle  \sigma _{ab}\rangle=&&\int_{0 }^{\infty}\tilde{C}^{(2)}_{a\neq b}(\tau){\rm d}\tau\\
\equiv&&T_aT_b\int_{0 }^{\infty}\Pi _{e;ab}(\tau ){\rm d}\tau\nonumber
\end{eqnarray}
Note, that there are no Ericson flcutuations in the $S$-matrix correlations. They show up only in the cross-sections, as outlined later.
\subsubsection{Elastic Autocorrelations}
The elastic case is derived in  close similarity to the inelastic one. In distinction to the latter, the average $S$-matrix element $\langle S^{res}_{aa}(E)\rangle $ is non-vanishing. This introduces a characteristic dependence on the level correlations. Proceeding as previously in the derivation of Eq.~(\ref{Eq:2pointcorababenergy}) gives, unless the transmission coefficient $T_{a}$  is close to unity,
\begin{eqnarray}
C^{(2)}_{aa}( \epsilon)
=&&2\pi\left \langle{\Gamma _{a}^2\over i\epsilon +\Gamma }\right\rangle\label{Eq:2-pointaabb}\\
-&&2\pi\int_{-\infty }^{\infty }{\rm d}r Y_2(r)\left\langle{\Gamma_{1a}\Gamma_{2a}\over i(\epsilon -r)+(\Gamma _1+\Gamma _2)/2}\right\rangle,\nonumber
\end{eqnarray}
where $\Gamma_i=\sum_e\Gamma_{ie}$.
Note, that only the fluctuating parts of the $S$-matrix elements contribute.  
The two-level cluster function $Y_2(r)$  is defined and discussed in Appendix \ref{App:spacingfunctions}. The first term in Eq.~(\ref{Eq:2-pointaabb}), already occurring in Eq.~(\ref{Eq:2pointcorababenergy}), measures the correlation between different parts of the same broadened resonance. The second one is generated by two broadened resonances at a distance $r$ correlated via the two-point cluster function $Y_2(r)$. 

For the present discussion we choose the Dyson (GOE) two-point cluster function $Y_{2}(r)$; see, e.g., Ref.~\cite {MEH67,DYS63}, Eq.~(5.69). Its Fourier transform yields the form factor $b(\tau)$ with $b(\tau =0)=1$, which is given in Eq.~(\ref{Eq:Dyson}). The Fourier transform of $C^{(2)}_{aa}( \epsilon)$ in Eq.~(\ref {Eq:2-pointaabb}) is similar to that of the inelastic $S$-matrix autocorrelation function in Eq.~(\ref{Eq:2-pointFourier}), except for an additional term arising due to level correlations,
\begin{eqnarray}
\label{Eq:2-pointtimeaabbnonav} 
\tilde{C}^{(2)}_{aa}(\tau) =&&(2\pi )^2 
\Big\{\left\langle\Gamma_{a}^2\exp (-2\pi\Gamma\tau )\right\rangle\\ 
-&& b(\tau )\langle\Gamma_{1a}\exp (-\pi \Gamma_1\tau )\rangle \langle\Gamma _{2a}\exp (-\pi \Gamma _2 \tau )\rangle\Big\}.\nonumber
\end{eqnarray}
With the short-hand notation of Appendix~\ref{App:shorthand} it takes the form
\begin{eqnarray}
  \hspace{-0.3in}&& 
\tilde{C}^{(2)}_{aa}(\tau) = 
\label{Eq:2-pointtimeaabb}
T_a^2\left[ \Pi _{e;aa}(\tau ) - b(\tau) \Pi _{e;a}^2 (\tau /2)\right]. 
\end{eqnarray}
Here, an elastic enhancement factor $A(k_a)$ appears with the value $A(2)=3$ for the standard case of a Porter-Thomas distribution; see Appendix~\ref{App:shorthand}.  \\
 
\begin{figure}[ht!]
\includegraphics[width=\linewidth]{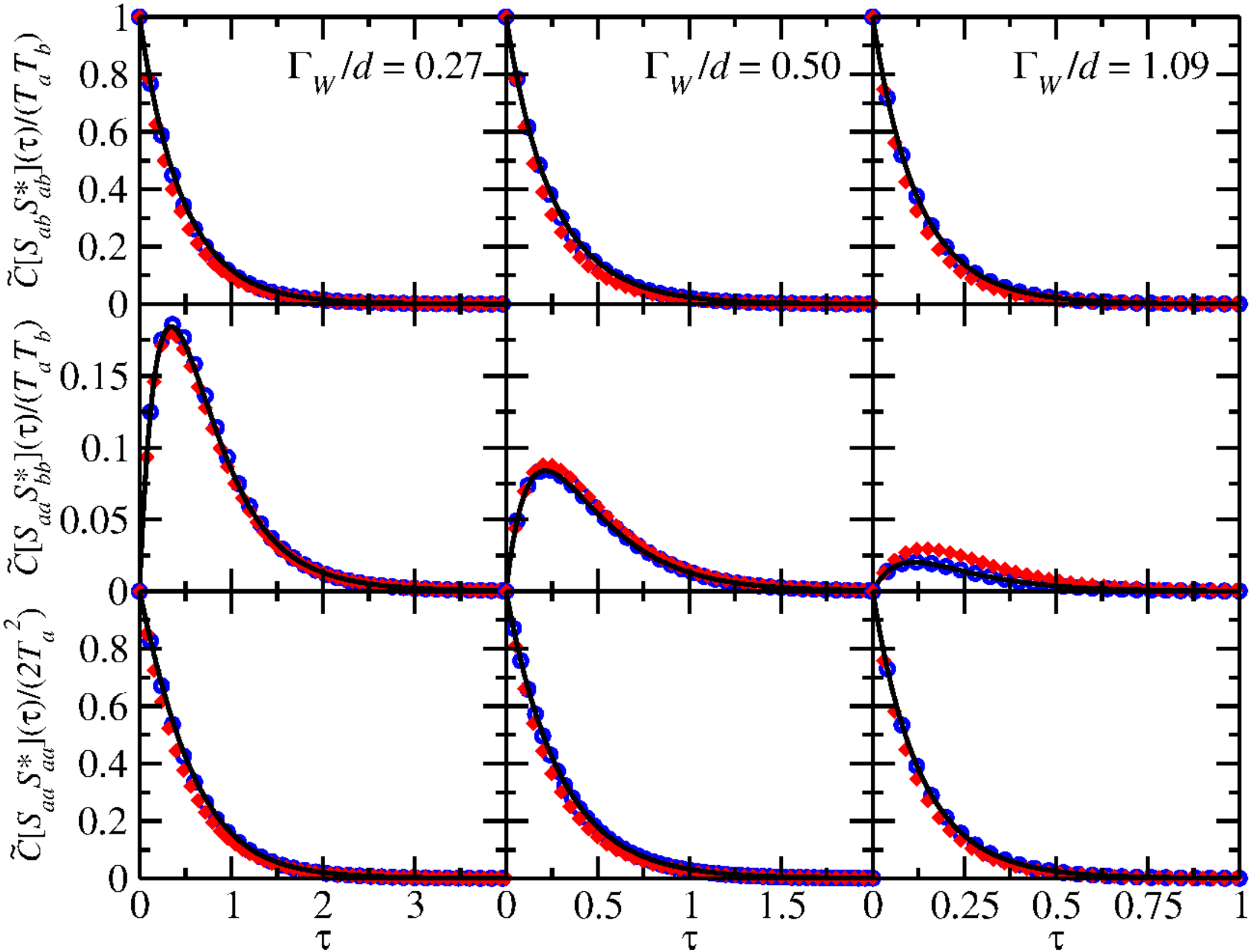}
\caption{$S$-matrix autocorrelation function in the time representation. The full black lines correspond to the analytical results, Eqs.~(24) and~(25) of Ref.~\cite{ERI13}, blue circles to those derived on the basis of the VWZ model~\cite{VER85} and red diamonds to the SWB model Eqs.~(\ref{Eq:PieabPorterThomas}) and~(\ref{Eq:2-pointtimeaabb}). The $\Lambda=52$ transmission coefficients corresponding to the values of $\Gamma_W/d$ in the insets are listed in~\reftab{table1}. All correlation functions were divided by $(1+\delta_{ab})T_aT_b$, predicted at  $\tau=0$ for the inelastic case ($b\ne a$) in the upper panels and the elastic one ($b=a$) in the lower panels. The cross-correlation functions ($b\ne a$) in the middle panels vanish there at $\tau=0$.}
\label{fig1}
\end{figure}

These results for the SBW model have been derived explicitly with the intention of applying them and their generalizations to studies of cross-section correlations for  chaotic or nearly chaotic systems. Since the results are approximate, it is essential to test the efficiency of the model by comparison to exact results. Our philosophy is that "the proof of the pudding is in the eating".   Based on RMT and the supersymmetry method, the VWZ approach~\cite{VER85} gives an exact analytical, although complex solution for the correlations of two $S$-matrix elements. Recently an accurate analytical approximation of the VWZ result has  been derived by one of us and compared to the SBW model in Ref.~\cite{ERI13}.  The VWZ results and their structure were very well reproduced for a variety of transmission coefficients and over many magnitudes of their size with the exception of $T_{a,b}\simeq 1$, corresponding to $\langle S_{aa}\rangle\simeq 0$. This is illustrated in Fig.~\ref{fig1}, where the analytic results deduced from Ref.~\cite{ERI13} are shown as full black lines and the VWZ ones are plotted as blue circles for diverse choices of the transmission coefficient. Their values were taken from experimental studies with microwave billiards~\cite{DIE10} and are listed in~\reftab{table1}. These curves are compared to the SBW model, shown as red diamonds in Fig.~\ref{fig1}. Only for the cross-correlation functions deviations of the SBW model from the analytical and the VWZ curves are visible. For $\tau =0$, all autocorrelation functions reproduce the predicted value $\tilde C^{(2)}_{ab}(\tau =0)=(1+\delta_{ab})T_aT_b$~\cite{GOR02}. This agreement between the different models gives confidence to the applicability of the SBW model to cross-section correlations in regions for which the VWZ approach provides no predictions.
 \section {Cross-Section Autocorrelation Functions\label{Sec:crosscorr}}
For simplicity of notation we limit the discussion mainly to the autocorrelation functions. 
\subsection {Derivation of the Irreducible Functions for the  Cross-Section Autocorrelations} 
The cross-section autocorrelation function in Eq.~(\ref{Eq:4pointcorrgeneralcrossA}) is approximated within the SBW model by replacing the $S$-matrix elements entering the two-, three- and four-point correlations in Eqs.~(\ref{Eq:2pointSA}),~(\ref{Eq:fluctuation}), and~(\ref{Eq:4pointfl}) by the resonant Breit-Wigner terms $S_{ab}^{res}$ given in Eq.~(\ref{Eq:S-abres}). This does not introduce additional loss of information, since constant terms such as $\langle S^{res}_{ab}\rangle $ do not contribute to $S^{fl}_{ab}$. Furthermore, the average diagonal matrix elements $\langle S_{aa}\rangle $ are taken to have the values which by unitarity define the corresponding transmission coefficients~\cite{footnote}. 

The $S$-matrix autocorrelation functions $C_{ab}^{(2)}(\epsilon )$ occurring in the expressions~(\ref{Eq:2pointSA}) and~(\ref{Eq:fluctuation}) have been evaluated in the previous section,~\refsec{Sec:2point}. Therefore, only the three-point function in the curly brackets in~(\ref{Eq:2pointSA}) and the irreducible four-point term~(\ref{Eq:4pointfl}) remain to be evaluated in the SBW model. The procedure is illustrated schematically for the latter, i.e., for the four-point $S$-matrix term $C_{ab}^{(4)}(\epsilon)$ defined in Eq.~(\ref{Eq:C^4(epsilon)}). It becomes in terms of the resonance amplitudes Eq.~(\ref{Eq:S-abres})
\begin{eqnarray}
\label{Eq:4pointcorrfunction}
C_{ab}^{(4)}(\epsilon)&&= \\
\Big\langle
\sum_{klmn}
&&\left[{\gamma _{ak}\gamma _{kb}\over (E-\epsilon /2- e_k )}\right]^{fl}
\left[{\gamma _{al}\gamma _{lb}\over (E-\epsilon /2- e_l^* )}\right]^{fl}
\times\nonumber\\
&&\left[{\gamma _{am}\gamma _{mb}\over (E+\epsilon /2- e_m }\right]^{fl}  
\left[ {\gamma _{an}\gamma _{nb}\over (E+\epsilon /2- e^{*}_n )} \right ]^{fl}  \Big\rangle\nonumber\\ 
-&&\left\vert\left\langle
\sum _{kl}\left[ {\gamma _{ak}\gamma _{kb}\over (E- e_k )}\right]^{fl}  \left[ {\gamma _{al}\gamma _{lb}\over (E- e^{*}_l )}\right]^{fl} \right \rangle \right |^2. \nonumber 
 \end{eqnarray}
The expression~(\ref{Eq:4pointfl}) depends only on the average of the four correlated $S$-matrix elements and is irreducible. The sum over the 4 resonant indices $k,l,m,n$ can be decomposed into terms with less summations by simply setting part of the indices equal. Here, however, it is important to ensure that contributions are not counted too often. We illustrate this for the self-correlation $(k=l=m=n)$ of the broadened state $k$ with itself. This joint contribution is denoted by $\mathcal {F}_4^{abab}(\epsilon)$ (the "diagonal term"~\cite{GOR02}). The sum over $k$ is counted twice, since one of the terms is compensated by the folded contribution from two broadened resonances with $(k=l)\ne (m=n)$.  This joint contribution is denoted by $\mathcal {F}_4^{abab}(\epsilon)$ (the "diagonal term"~\cite{GOR02}). Two additional counterterms are produced by the products of pairs occurring in the second term of Eq.~(\ref{Eq:4pointcorrfunction}), i.e., in the Ericson fluctuation term, and in the pairs with poles in the same half of the complex energy plane. These linked terms correspond to $(k=n)\ne (l=m)$ and $(k=m)\ne (l=n)$, respectively, whereas the term $k=l=m=n$ is already counted in both cases. Their joint contribution is denoted by $\mathcal {G}_4^{abab}(\epsilon)$. Additional counterterms are generated by a single resonant contribution correlated to 3 other linked ones, i.e., counterterms corresponding to permutations of the type $k\neq (l=m=n)$.  They are combined in a term denoted as $\mathcal {H}_4^{abbb}(\epsilon)$. 
In principle, Eq.~(\ref{Eq:4pointcorrfunction}) depends explicitly on higher order  level cluster terms denoted by $Y_n(r)$. We conjecture that contributions from terms involving irreducible three-point (n=3) and four-point (n=4) level cluster functions are negligible, since the associated levels mutually repel each other and correspond to uncorrelated amplitudes.

 The procedures presented here qualitatively are the basis for the results given in~\refsec{Sec:irreducible} as well as for those given in Appendix~\ref{App:Results}.

\subsection{Irreducible Functions for the Cross-Section Correlations\label{Sec:irreducible}}
The general cross-section correlation function $C_{ab}(\epsilon)$ is expressed by irreducible $S$-matrix correlation functions and can then be evaluated analytically as described above for the irreducible four-point functions~(\ref{Eq:4pointcorrfunction}) and following for the autocorrelation case with $(cd)=(ab)$. 
The procedure is basically a generalization of that for the correlation functions for two $S$-matrix elements. Analytical expressions are obtained for these functions by observing that the ensemble average over the resonance positions can, as for the two-point function, be performed as an energy average in the standard way. The result then follows using complex integration which gives it in terms of the residues at the complex poles. The results are given for the general case with four observed channels $(abcd)$ and  presented separately for the energy and time representations. \\
 %%%%%%%%%%%%%%%%%%%%%%%%%%%%%%%%%%%
\subsubsection{The Energy Representation}
 %%%%%%%%%%%%%%%%%%%%%%%%%%%%%%%%%%%
 The cross-section autocorrelation function $C_{ab}(\epsilon)$ in Eq.~(\ref{Eq:4pointcorrgeneralcrossA}) of the observed channels $a,b,c,d$ is expressed in terms of irreducible four-point functions $\mathcal {F}_4^{abab}(\epsilon),\mathcal {G}_4^{abab}(\epsilon),\mathcal {H}_4^{abab}(\epsilon)$ corresponding to three irreducible terms of Eq.~(\ref{Eq:4pointfl}), the irreducible three-point function $C_{aa}^{(3)}(\epsilon)$ defined in~(\ref{Eq:2pointSA}), and the two-point functions entering~(\ref{Eq:2pointSA}) and~(\ref{Eq:fluctuation}). The latter are defined fully by the explicit $S$-matrix autocorrelation functions given in~\refsec{Sec:2point}. The inelastic two-point cross-section correlation function reads 
\begin{eqnarray}
C_{ab}(\epsilon) 
\label{Eq:fluctuationAApp}
=&&| C^{(2)}_{ab}(\epsilon  )|^2\\    
+&&\mathcal {F}_4^{abab}(\epsilon) +\mathcal {G}_4^{abab}(\epsilon)\label{Tabenergy2} 
\end{eqnarray}
and the elastic one
\begin{eqnarray}
&&C_{aa}(\epsilon)=\label{Eq:Cn=2elastic}\\
&&\mathcal {F}_4^{aaaa}(\epsilon)+2\Re\left\{ (1-\langle S_{aa}\rangle)^2C^{(2)}_{aa}(\epsilon )\right\}\label{Taaenergy1}\\
+&&\mathcal {G}_4^{aaaa}(\epsilon)+\left|C^{(2)}_{aa}(\epsilon )\right|^2\label{Taaenergy2}\\
+&&\mathcal {H}_4^{aaaa}(\epsilon)+\Re\mathcal{F}_3^{aaaa}(\epsilon).\label{Taaenergy3} 
\end{eqnarray}
Here, $\Re\mathcal{F}_3^{aaaa}(\epsilon)$ denotes a three-point correation function defined in terms of that given in~(\ref{Eq:3pointflA}) as
\begin{equation}
\Re\mathcal{F}_3^{aaaa}(\epsilon)=2(1-\langle S_{aa}\rangle)\Re\left[C_{aa}^{(3)}(\epsilon)+C_{aa}^{(3)}(-\epsilon)\right].
\end{equation}
The explicit analytical expressions for the irreducible $S$-matrix correlation functions are determined as described in the last part of the previous subsection. The different contributions to the correlation functions are even in the energy increment $\epsilon$. In this article we mainly restrict the discussions to $S$-matrix and cross-section autocorrelations. Then the indices $c,d$ take the values $a$ or $b$ only.

The terms with $(k=l=m=n)$ and $(k=l) \neq (m=n)$ in Eq.~(\ref{Eq:4pointcorrfunction}) yield
\begin{eqnarray}
&&\mathcal{F}_4^{abcd}(\epsilon) = \label{Eq:F_4App} 
4\pi\left\langle\frac{\Gamma_{1a}\Gamma_{1b}\Gamma_{1c}\Gamma_{1d}}{\Gamma_1^2}\frac{\Gamma_1}{\epsilon ^2+\Gamma _1^2}\right\rangle\\   
&&-2\pi\int drY_2(r) 
\left\langle\frac{ \Gamma _{1a}\Gamma _{1b}\Gamma _{2c}\Gamma _{2d}}{\Gamma _1\Gamma _2}{(\Gamma _1+\Gamma _2)\over (r+\epsilon  )^2+{1\over 4}(\Gamma _1+\Gamma _2) ^2 } \right\rangle,\nonumber
\end{eqnarray}
while the terms with $(k=m) \neq (l=n)$ and $(k=n) \neq (l=m) $ in Eq.~(\ref{Eq:4pointcorrfunction}) give
\begin{eqnarray}
&&\mathcal {G}_4^{abcd}(\epsilon) = 
\label{Eq:G_4App}-2\pi\int_{-\infty }^{\infty }{\rm d}rY_2(r)\times\\
&&\nonumber\Big\langle{\Gamma _{1a}\Gamma _{1b}\Gamma _{2c}\Gamma _{2d}
\over r^2+{1\over 4}(\Gamma _1+\Gamma _2) ^2} 
\Big\{ {\Gamma _1 \over \epsilon ^2 +\Gamma _1 ^{2}}+{\Gamma _2\over \epsilon ^2 +\Gamma _2 ^{2}}\\  
&&\nonumber -{(\Gamma _1+\Gamma _2)\over (r-\epsilon )^2+{1\over 4}(\Gamma _1+\Gamma _2) ^{2}} 
\left(3 -{2r\epsilon +(\Gamma _1+\Gamma _2)^2 \over r^2 +{1\over 4}(\Gamma _1+\Gamma _2)^2 }\right)\Big\}\Big\rangle .\nonumber
\end{eqnarray}
The irreducible function $\mathcal {G}_4^{abcd}$ is non-vanishing only for indices $(ab;cd)$ referring to the combinations $(aa;aa)$, $(ab;ab)$ and $(aa;bb)$.

Terms of the type $k\neq  (l=m=n)$ in Eq.~(\ref{Eq:4pointcorrfunction}) lead to
\begin{eqnarray}
&&\mathcal {H}_4^{abcd}(\epsilon) = \label{Eq:H_4App} 
-2\pi\delta_{ab }\int_{-\infty}^{\infty}{\rm d}rY_{2}(r)\times\\
&&\nonumber\Big\langle\frac{\Gamma_{1a }\Gamma_{2b}\Gamma_{2c}\Gamma_{2 d }}{\Gamma_2(r^2+{1\over 4}(\Gamma _1+\Gamma _2) ^2)}\times\\
&&\nonumber\Big\{\frac{(\Gamma _1+\Gamma _2) \Gamma _2}{\epsilon ^2 +\Gamma _2 ^{2}}
-2{(r^2-\epsilon r)-{1\over 4}(\Gamma _1+\Gamma _2) ^2\over (r-\epsilon)^2 +{1\over 4}(\Gamma _1+\Gamma _2) ^2 }\Big\}\Big\rangle\\
&&\nonumber +(ab )\leftrightarrow (cd). 
\end{eqnarray}
In addition the reducible terms corresponding to the autocorrelations in Eq.~(\ref{Eq:4pointcorrfunction})  generate a three-point correlation function $C_{ab}^{(3)}$ for the elastic case. Using the definition of the three-point function from Eq.~(\ref{Eq:3pointflA}) gives
\begin{eqnarray}
&&\label{Eq:F_3App}\mathcal {F}_3^{abcd}(\epsilon )=4\pi\delta_{ab }\left(1-\langle S_{aa }\rangle\right)\times\\
&&\nonumber\Big\{ \Big\langle\frac{\Gamma_{1a}\Gamma_{1c}\Gamma_{1 d }}{\epsilon ^2+\Gamma ^2_1}\Big\rangle\\ &&\nonumber
-\int_{-\infty}^{\infty}{\rm d}rY_2(r)\Big\langle\frac{\Gamma_{1a}\Gamma_{2c} \Gamma_{2 d }}{\Gamma _2}\frac{\frac{1}{2}(\Gamma _1+\Gamma _2)}{(r-\epsilon )^2+{1\over 4}(\Gamma _1+\Gamma _2)^2 }\Big\rangle \\
&&\nonumber -\delta_{ cd}\int_{-\infty}^{\infty}{\rm d}rY_2(r) 
\Big\langle\frac{\Gamma _ {1a} \Gamma _{1c} \Gamma _{2d}}{r^2+\frac{1}{4}(\Gamma_1+\Gamma_2)^2}\times\\ &&\nonumber
\left[\frac{\frac{1}{2}(\Gamma _1+\Gamma _2)\Gamma _1}{\epsilon^2 +\Gamma _1^2}-\frac{r^2-\epsilon r-\frac{1}{4}(\Gamma _1+\Gamma _2)^2}{(r-\epsilon)^2 +{1\over 4}(\Gamma _1+\Gamma _2) ^2 }\right]\Big\rangle\Big\}\\ &&\nonumber+(ab )\leftrightarrow (cd).
\end{eqnarray}
Additional reducible contributions are generated by the reduction into $S$-matrix autocorrelation functions.  The terms $ C^{(2)}_{aa}(\epsilon) $ in Eq.~(\ref{Eq:2pointSA}) and $|C^{(2)}_{ab}(\epsilon)|^2 $ in Eq.~(\ref{Eq:fluctuation})  are obtained from Eqs.~(\ref{Eq:2pointcorababenergy}) and~(\ref{Eq:2-pointaabb}). For the numerical evaluation of the cross-section autocorrelation function~(\ref{Eq:Cn=2elastic}) we used the analytical results of Ref.~\cite{ERI13} for $C^{(2)}_{ab}(\epsilon)$. 
\subsubsection{The Time Representation\label{Sec:Timecorr}}
The cross-section autocorrelation function in the time representation, $\tilde {C}_{ab}(\tau ) $, is obtained from the Fourier transforms of the functions in Eqs.~(\ref{Eq:F_4App}) -~(\ref{Eq:F_3App}); see  Appendix~\ref{Sec:Timecorrelations}. The inelastic autocorrelations are given by 
\begin{eqnarray}
&&\tilde {C}_{ab}(\tau )=\label{Eq:CabtimeApp}\\
&&\int_0^{\infty}{\rm d}\lambda  \tilde {C}^{(2)}_{ab}(\lambda ) \tilde {C}^{(2)}_{ab}(\lambda +|\tau |)\label{Tabtime1}\\
&&+\tilde {\mathcal{F}}_4^{abab}(\tau )+\tilde {\mathcal{G}}_4^{abab}(\tau ),\label{Tabtime2}
\end{eqnarray}
the elastic ones by 
\begin{eqnarray}
&&\tilde {C}_{aa}(\tau )=\label{Eq:CaatimeApp}~~~~~\\
&&\tilde {\mathcal{F}}_4^{aaaa}(\tau )
+2\Re\left\{(1-\langle S_{aa} \rangle)^2\tilde{C}^{(2)}_{aa}(\tau )\right\}\label{Taatime1}~~~~~\\
+&&\tilde{\mathcal{G}}_4^{aaaa}(\tau )
+\int_0^{\infty}{\rm d}\lambda\tilde {C}^{(2)}_{aa}(\lambda ) \tilde {C}^{(2)}_{aa}(\lambda +|\tau | )\label{Taatime2}~~~~~\\
+&&\tilde {\mathcal{H}}_4^{aaaa}(\tau )+\Re\tilde{\mathcal{F}}_3^{aaaa}(\tau ).~~~~~\label{Taatime3}
\end{eqnarray}
Using the short-hand notation of Eqs.~(\ref{Eq:Pi_eabc}) and~(\ref{Eq:shorthandsumgeneral}) the separable expression $\tilde{\mathcal{F}}_4^{abcd}(\tau )$ is obtained from Eq.~(\ref{Eq:F_4App}), 
\begin{eqnarray}
&&\tilde{\mathcal{F}}_4^{abcd}(\tau ) =\label{Eq:F_4timeApp}T_aT_bT_cT_d\times \\
&&\nonumber
 \Big\{\int_0^{\infty}\lambda {\rm d}\lambda \Pi _{e;abcd}( \lambda +| \tau |)\\ &&\nonumber
-b(\tau)\int_0^{\infty }\int_0^{\infty } {\rm d}\mu {\rm d}\lambda   \Pi _{e;ab}( \mu +|\tau |/2) \Pi _{e;cd}( \lambda  +|\tau |/2)\Big\}.\nonumber 
\end{eqnarray}
Similarly, the irreducible functions $\tilde {\mathcal{G}}_4^{abcd} (\tau )$, $\tilde {\mathcal{H}}_4^{abcd}(\tau )$ and $\tilde {\mathcal{F}}_3^{abcd}(\tau )$ are derived from Eqs.~(\ref{Eq:G_4App})-(\ref{Eq:F_3App}), yielding
\begin{eqnarray}
&&\tilde{\mathcal{G}}_4^{abcd}(\tau )=\label{Eq:G_4timApp}
-  T _{a}^2T _{b}^2\times\\ &&\nonumber
 \Big\{\int_0^{\infty }\int _0^{\infty } {\rm d}\mu 
{\rm d}\lambda  b(\lambda  )
 \Pi _{e;ab} ((\mu+\lambda )/2)\Pi _{e;\gamma  \delta } ((\mu+\lambda )/2 +|\tau|)\\ &&\nonumber+\int_{0}^{\infty} 
\lambda d\lambda  b(\lambda +|\tau | ) \Pi _{e;ab} ((\lambda +|\tau |)/2 )\Pi _{e;\gamma  \delta} ((\lambda +|\tau| )/2 ) 
 \Big\}, 
 \end{eqnarray}

 \begin{eqnarray}
&&\tilde{\mathcal{H}}_4^{abcd} (\tau ) = \label{Eq:H_4timeApp}
    - \delta _{ab}    T _{a }T _{b}T _{c}T_{d }\times\\ 
&&\nonumber    
\Big\{\int_{|\tau |} ^{\infty} {\rm d}\mu   \int_0^{\infty }{\rm d}\lambda   b(\lambda) \Pi _{e;a } (\lambda /2)\Pi _{e;bcd}  [\mu +\lambda /2]\\ &&\nonumber
+\int_0^{\infty} {\rm d}\mu    \int_{|\tau |}^{\infty }{\rm d}\lambda  b(\lambda )\Pi _{e;a }  [\lambda /2]~ \Pi _{e;b cd } [\mu +\lambda /2] 
\Big\}\\
&& \nonumber +(ab) \leftrightarrow  (cd), 
\end{eqnarray}
and
\begin{eqnarray}
&&\tilde{\mathcal{F}}_3^{abcd} (\tau )
 = \label{Eq:F_3timeApp}
 \left(1-\langle S_{aa} \rangle\right)  \delta _{ab}   T _{a}T _{c}T_{d} \times \\
&&\nonumber
\Big\{\int_0^{\infty}{\rm d}\lambda \Pi_{e;acd} (\lambda +|\tau |))   -  b(\tau)  \Pi_{e;a} (| \tau |/2)\times\\ &&\nonumber 
   \int_0^{\infty}{\rm d}\lambda \Pi _{e;cd}(\lambda +|\tau |/2) \\ &&\nonumber
- \delta _{cd} \Big[  \int_0^{\infty }{\rm d}\lambda  b(\lambda )  \Pi _{e;d}(\lambda /2)   \Pi _{e;ac}(\lambda /2+|\tau |)\\ &&\nonumber 
+\int_{|\tau |}^{\infty }{\rm d}\lambda  b(\lambda )\Pi _{e;ac} ( \lambda /2) \Pi_{e;d}  ( \lambda  /2)   \Big]  \Big\} \nonumber
 \\
 && +(ab)\leftrightarrow (cd). \nonumber
\end{eqnarray}
As in the energy representation, the function $\tilde{\mathcal{G}}_4^{abcd} (\tau )$ has non-vanishing contributions only for indices $(ab;cd)$ equal to $(aa;aa)$, $(ab;ab)$ or $(aa;bb)$, whereas the function $\tilde{\mathcal{H}}_4^{abcd} (\tau )$ contributes only when either $a$ and $b$ or $c$ and $d$ coincide. 

The energy and the time representations of the cross-section autocorrelation functions are complementary and give different insights. The separable property in the time representation gives contributions from the individual channels $e$ in a multiplicative form, contrary to the folding in the energy representation. This property allows the detailed description of the cross-section correlations in terms of the transmission coefficients $T_e$ of the individual channels $e$. Furthermore it simplifies considerably the numerical evaluation of the cross-section correlation functions. The more general form in Eqs.~(\ref{Eq:F_4App})-(\ref{Eq:F_3App}) is useful in the limit of constant total widths. This latter case gives a global overall view of the 'forest' of all the channels and their net effect, while the separable description in Eqs.~(\ref{Eq:F_4timeApp}) -~(\ref{Eq:F_3timeApp}) emphasizes the effects of the individual channels, the 'trees in the forest'. 

\begin{figure}[ht!]
        \includegraphics[width=0.9\linewidth]{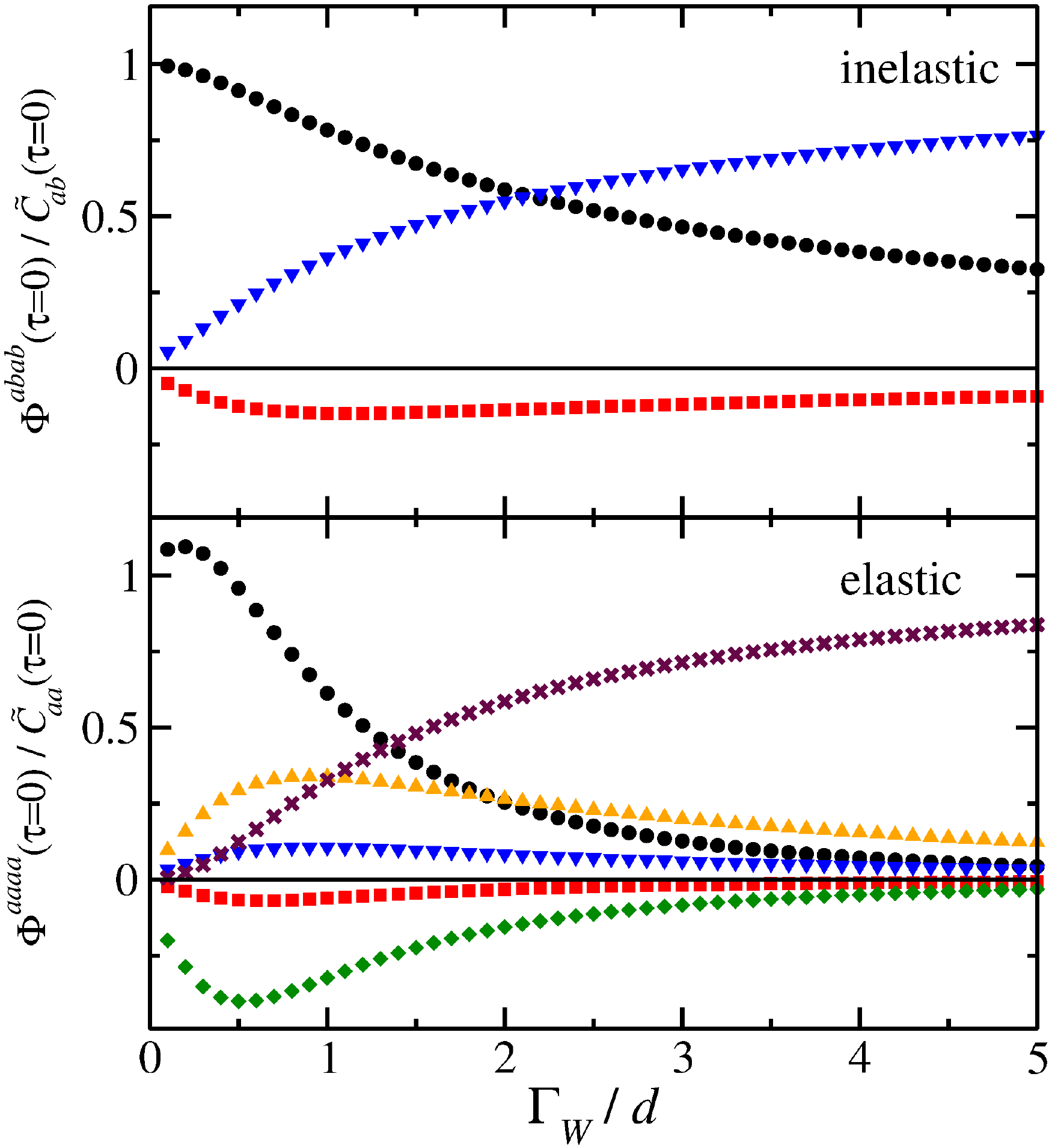}
        \caption{Relative contributions of the individual terms entering Eqs.~(\ref{Eq:CabtimeApp}) and~(\ref{Eq:CaatimeApp}) to the cross-section correlation functions $\tilde C_{ab}(\tau=0)$ versus $\Gamma_W/d$ for $52$ equal transmission coefficients. Black dots correspond to $\Phi=\tilde{\mathcal{F}}_4$, red squares to $\Phi=\tilde{\mathcal{G}}_4$, green diamonds to $\Phi=\tilde{\mathcal{H}}_4$, orange triangles-up to $\Phi=\tilde{\mathcal{F}}_3$. Blue triangles-down display the Fourier transform of the Ericson fluctuation term in Eq.~(\ref{Tabtime1}), maroon crosses that of the second term in Eq.~(\ref{Taatime1}). Upper panel: the inelastic case $a=1,\, b=2$. Lower panel: the elastic case $a=b=1$.}
\label{fig4}
\end{figure}

Figure~\ref{fig4} shows for $\tau=0$ the fractions of the contributions of the functions entering~(\ref{Eq:CabtimeApp}) and~(\ref{Eq:CaatimeApp}) to the cross-section correlation function $\tilde C_{ab}(\tau)$ versus $\Gamma_W/d$. These results are qualitatively characteristic of the relative sizes of the corresponding terms for $\tau\ne 0$. In the inelastic case (upper panel) the cross-section term $\tilde C_{ab}(0)$ is well approximated by $\tilde{\mathcal{F}}_4^{abab}(0)$ (black dots) for $\Gamma_W/d\lesssim 2$. For larger values of $\Gamma_W/d$ the Fourier transform of the Ericson fluctuation term~(\ref{Tabtime1}), shown as blue triangles-down, becomes dominant. Thus, in the inelastic case, $\tilde {C}_{ab}(\tau )$ approaches the function 
\begin{equation}
\label{Eq:Cabtimeas}
\tilde {C}_{a\ne b}^{(as)} (\tau )=\int_0^{\infty} {\rm d}\lambda  \tilde {C}^{(2)}_{ab}(\lambda ) \tilde {C}^{(2)}_{ab}(\lambda +|\tau |)
\end{equation}
for large values of $\Gamma_W/d$. 

\begin{figure}[ht!]
\includegraphics[width=\linewidth]{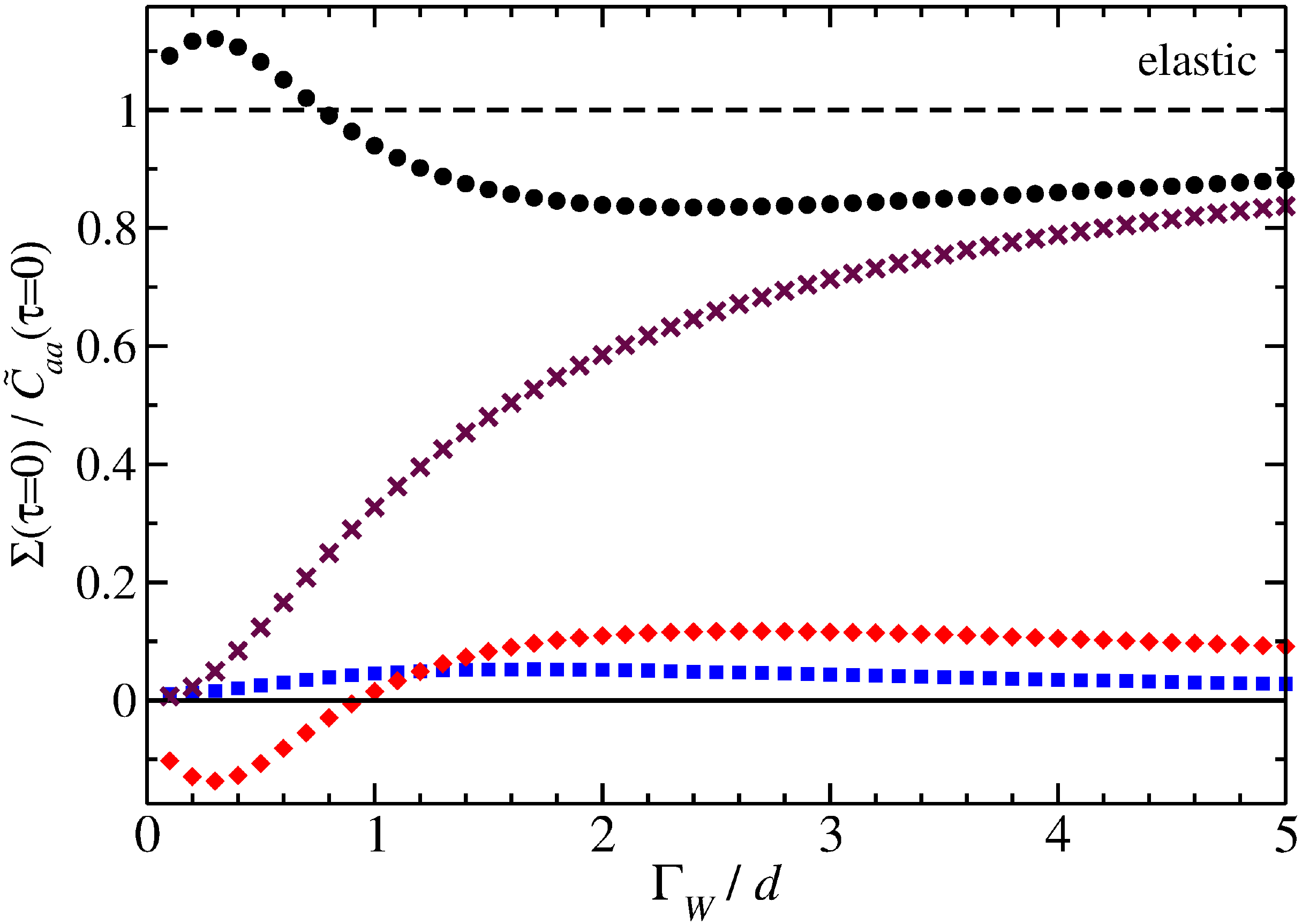}
\caption{Same as Fig.~\ref{fig4}, for the specific combinations $\Sigma$ of the individual contributions given in (\ref{Taatime1}) [black dots], (\ref{Taatime2}) [blue squares], and (\ref{Taatime3}) [red diamonds], respectively. It is clearly visible that the cross-section correlation function is well approximated by the sum Eq.~(\ref{Taatime1}) of the two terms which dominate the regions of isolated and overlapping resonances, respectively. With increasing $\Gamma_W/d$, $\tilde C_{aa}(\tau=0)$ approaches $\tilde C_{aa}^{(as)}(\tau =0)$ given in (\ref{Eq:Caatimeas}), plotted as maroon crosses.}
\label{fig4a}
\end{figure}

In the elastic case, the contribution from $\tilde{\mathcal{F}}_4^{aaaa}(0)$ (black dots) is dominant for small $\Gamma_W/d$ due to self correlations, like in the inelastic one. The three-point function $\tilde{\mathcal{F}}_3^{aaaa}(0)$ (orange triangle-up) and $\vert\tilde{\mathcal{H}}_4^{aaaa}(0)\vert$ (green diamonds) also have comparatively large values for $\Gamma_W/d\lesssim 1.5$. However, these terms cancel each other systematically to a large degree as illustrated in Fig.~\ref{fig4a} (red diamonds). The two-point correlation function~(\ref{Taatime1}) becomes dominant for $\Gamma_W/d\gtrsim 1.5$. Contrary to the inelastic case, the Ericson fluctuation term (blue triangles-down) and also $\tilde{\mathcal{G}}_4^{aaaa}(0)$ (red squares) are small for all values of $\Gamma_W/d$ and the sum of these terms of opposite signs is vanishingly small, as illustrated in Fig.~\ref{fig4a} (blue squares). As a consequence, $\tilde C_{aa}(\tau)$ is well approximated by the sum of the two terms in~(\ref{Taatime1}), shown as black dots in Fig.~\ref{fig4a}. With increasing $\Gamma_W/d$ the function $\tilde{\mathcal{F}}_4^{aaaa}(0)$ becomes vanishingly small so that for large $\Gamma_W/d$ the function $\tilde {C}_{aa}(\tau )$ is proportional to the Fourier transform of the $S$-matrix autocorrelation function,
\begin{equation}
\tilde {C}_{aa}^{(as)}(\tau )=\label{Eq:Caatimeas}
2\Re\left\{(1-\langle S_{aa}\rangle)^2\tilde{C}^{(2)}_{aa}(\tau )\right\},
\end{equation}
which is plotted as maroon crosses in Fig.~\ref{fig4a}
In the energy representation the functions entering Eqs.~(\ref{Eq:fluctuationAApp}) and~(\ref{Eq:Cn=2elastic}) exhibit the same relative behavior as in the time representation. In the inelastic case, the cross-section correlation function approaches with increasing $\Gamma_W/d$ the Ericson fluctuation term 
\begin{equation}
C_{a\ne b}^{(as)}(\epsilon)=\label{Eq:Cabenergyas}
|C^{(2)}_{ab}(\epsilon)|^2,    
\end{equation}
and in the elastic one twice the $S$-matrix autocorrelation function
\begin{equation}
C_{aa}^{(as)}(\epsilon)=\label{Eq:Caaenergyas}
2\Re\left\{(1-\langle S_{aa}\rangle)^2C^{(2)}_{aa}(\epsilon )\right\}.
\end{equation}
These results are in accordance with those obtained in Refs.~\cite{ERI60,ERI63}, but now with a well-defined range of validity.  
The conclusions to be drawn from the above observations can be summarized as follows: in the limit of narrow resonances the diagonal term $\mathcal{F}_4^{abab}$ and its Fourier transform are strongly dominant. This behavior reflects the self-correlation of a narrow resonance with itself. It is particularly apparent in the energy representation in Eq.~(\ref{Eq:F_4App}), since level repulsion suppresses the contribution from other resonances for $\epsilon =0$. In the time representation a correlation hole is produced by level repulsion for small values of $\tau$. With increasing total width the r\^ole of the diagonal term diminishes and the most reducible terms in Eqs.~(\ref{Eq:fluctuationAApp}) and~(\ref{Eq:Cn=2elastic}) and, similarly, in Eqs.~(\ref{Eq:CabtimeApp}) and~(\ref{Eq:CaatimeApp}) become the dominant ones. The SBW model strongly suggests that the asymptotic region begins appoximately at $\Gamma_W/d \simeq\pi$. Above this value the Ericson fluctuation term and the linear contribution from the $S$-matrix autocorrelation function rapidly become the prevailing terms in the inelastic and the elastic case, respectively.
\section{Experimental, Analytical and Numerical Tests\label{Exp}}
To test the SBW model we used data from experiments with microwave resonators, known exact analytical results at the energy increment $\epsilon =0$ and we performed RMT simulations similar to those presented in Refs.~\cite{Dietz2009a,DIE10}. 
\subsection{Experimental Details}
For the experimental test we used the same data as in Ref.~\cite{DIE10}. There, a chaotic scattering system was realized with a flat microwave resonator with the shape of a tilted-stadium billiard~\cite{Dietz2009a,DIE10,DIE10A}. The dynamics of the corresponding classical billiard is chaotic~\cite{Primack1994}. Thus, according to the Bohigas-Giannoni-Schmit conjecture~\cite{Bohigas1984}, the fluctuation properties of the eigenvalues of the associated quantum billiard are described by random matrices from the GOE~\cite{MEH67}. The modes in the resonator were coupled to the exterior via two antennas. For the determination of the $S$-matrix elements $S_{ab}$ describing the scattering process from antenna $b$ to antenna $a$ with $a,\ b\in\{1,\ 2\}$, a vector network analyzer coupled microwave power into the resonator via antenna $b$ and determined the relative phase and amplitude of the transmitted signal at antenna $a$. Resonance spectra were measured with a step size $\Delta f=$100kHz in the frequency range 5-25~GHz. To ensure that averages of the resonance widths and the resonance spacings were approximately constant, we analyzed the spectra in 1~GHz frequency intervals, yielding $10^4$ data points each. Furthermore a movable scatterer was inserted into the microwave resonator to gather in each frequency interval 8 independent data sets. At low excitation frequencies the resonances are isolated, i.e., the mean resonance widths are small compared to the resonance spacing $d$. There  the number of resonances in a 1~GHz window is comparatively small. With increasing excitation frequency the resonances begin to overlap. We used the Weisskopf formula~\cite{Blatt1952,Dietz2011}
\begin{equation}
        2\,\pi\, \frac{\Gamma_W}{d} = \sum_c T_c = T_1 + T_2 +
        \tau_{\rm abs} 
        \label{eqn:GammaD}
\end{equation}
to characterize the frequency intervals. Here, $T_1$ and $T_2$ are the transmission coefficients corresponding to the antennas. They are determined from the measured reflection spectra using Eq.~(\ref{transm}). Ohmic absorption in the walls of the resonator was modeled by a large number of fictitious channels. The sum of the corresponding transmission coefficients yields $\tau_{\rm abs}$. Its value was determined from a fit of the analytic result for the $S$-matrix autocorrelation function obtained from the VWZ model~\cite{VER85} to the experimental one. The values for the transmission coefficients and $\Gamma_W/d$ in each 1 GHz window are listed in~\reftab{table1}. Note that, generally, $\tau_{abs}\gg T_1,\, T_2$. The reason for this is that losses due to absorption in the walls dominate those through the antennas, since the coupling of the antenna states to the resonator modes is only weak. Detailed information on the experiments, the analysis of the experimental data and the systematic and statistical errors can be found in Refs.~\cite{DIE08,DIE10A}.
\begin{table}[b]
\caption{The values of the transmission coefficients $T_1,\, T_2$ and of $\tau_{abs}=\sum_{i=3}^\Lambda T_i$ with equal transmission coefficients $T_3=...=T_\Lambda$, the Weisskopf estimate $\Gamma_W/d$ for the total widths and the average ratio $\overline{\rho}_{1,2}=(\langle\Gamma_1\rangle+\langle\Gamma_2\rangle )/(2\Gamma_W)$ for the experimental spectra in the 1~GHZ frequency windows. The numerical simulations associated with these data were all performed with $\Lambda =52$.}
\label{table1}
\begin{tabular}{cccccc}
\hline
\hline
 ~GHz~ & ~$T_1$~ & ~$T_2$~& $\tau_{abs}$ & $\Gamma_W/d$ &$\overline{\rho}_{1,2}$\\
\hline
5 - 6 & 0.012 & 0.014 & 0.331 & 0.06 & 0.036\\
6 - 7 & 0.031 & 0.032 & 0.462 & 0.08 & 0.060\\
7 - 8 & 0.037 & 0.039 & 0.588 & 0.11 & 0.057\\
8 - 9 & 0.079 & 0.067 & 0.728 & 0.14 & 0.083\\
9 -10 & 0.097 & 0.130 & 0.810 & 0.17 & 0.109\\
10-11 & 0.178 & 0.222 & 1.011 & 0.23 & 0.142\\
11-12 & 0.256 & 0.233 & 1.205 & 0.27 & 0.144\\
12-13 & 0.303 & 0.327 & 1.288 & 0.31 & 0.164\\
13-14 & 0.401 & 0.415 & 1.546 & 0.38 & 0.173\\
14-15 & 0.332 & 0.379 & 1.793 & 0.40 & 0.142\\
15-16 & 0.455 & 0.353 & 1.891 & 0.43 & 0.150\\
16-17 & 0.399 & 0.404 & 2.046 & 0.45 & 0.141\\
17-18 & 0.417 & 0.475 & 2.274 & 0.50 & 0.141\\
18-19 & 0.528 & 0.496 & 2.598 & 0.58 & 0.141\\
19-20 & 0.480 & 0.457 & 3.265 & 0.67 & 0.111\\
20-21 & 0.583 & 0.538 & 4.135 & 0.84 & 0.107\\
21-22 & 0.638 & 0.558 & 4.739 & 0.94 & 0.100\\
22-23 & 0.710 & 0.593 & 4.806 & 0.97 & 0.107\\
23-24 & 0.784 & 0.665 & 4.903 & 1.01 & 0.114\\
24-25 & 0.694 & 0.796 & 5.344 & 1.09 & 0.109\\
\hline
\hline
\end{tabular}
\end{table}
\subsection{The RMT model}
To model chaotic scattering systems for values of $\Gamma_W/d$ larger than achieved in the experiments, we performed RMT simulations. For this we used the $S$-matrix formalism developed in~\cite{Mahaux1969} in the context of compound-nucleus reaction theory. The associated unitary $S$ matrix has the general form
\begin{equation}
        S(E) = \II - 2\pi i W \big(E\II-H + i\pi W^T W\big)^{-1} W^T\, .
        \label{eqn:Sab}
\end{equation}
Here, the Hamiltonian $H$ describes the internal dynamics. The matrix elements of $W$ specify the couplings of its states to the open channels~\cite{VER85,Dietz2007a}. In order to model chaotic, time-reversal invariant scattering systems like, e.g., the microwave resonator described above, we inserted in Eq.~(\ref{eqn:Sab}) for $H$  a real and symmetric random matrix from the GOE~\cite{MEH67}. The matrix entries of $W$ were chosen as real Gaussian-distributed random numbers with zero mean. The entries of the matrix $WW^T$~\cite{Mahaux1969,VER85} determine the transmission coefficients $T_e$, with $e=1,\cdots,\Lambda$. In the simulations, the number of open channels $\Lambda$, that is, the dimension of $S(E)$ was set to $52$, that of $H$ to $N=200$. The RMT results were obtained with an ensemble of $1000$ random scattering matrices. 
\subsection{Variances of the cross-sections}
\begin{figure}[ht!]
        \includegraphics[width=0.9\linewidth]{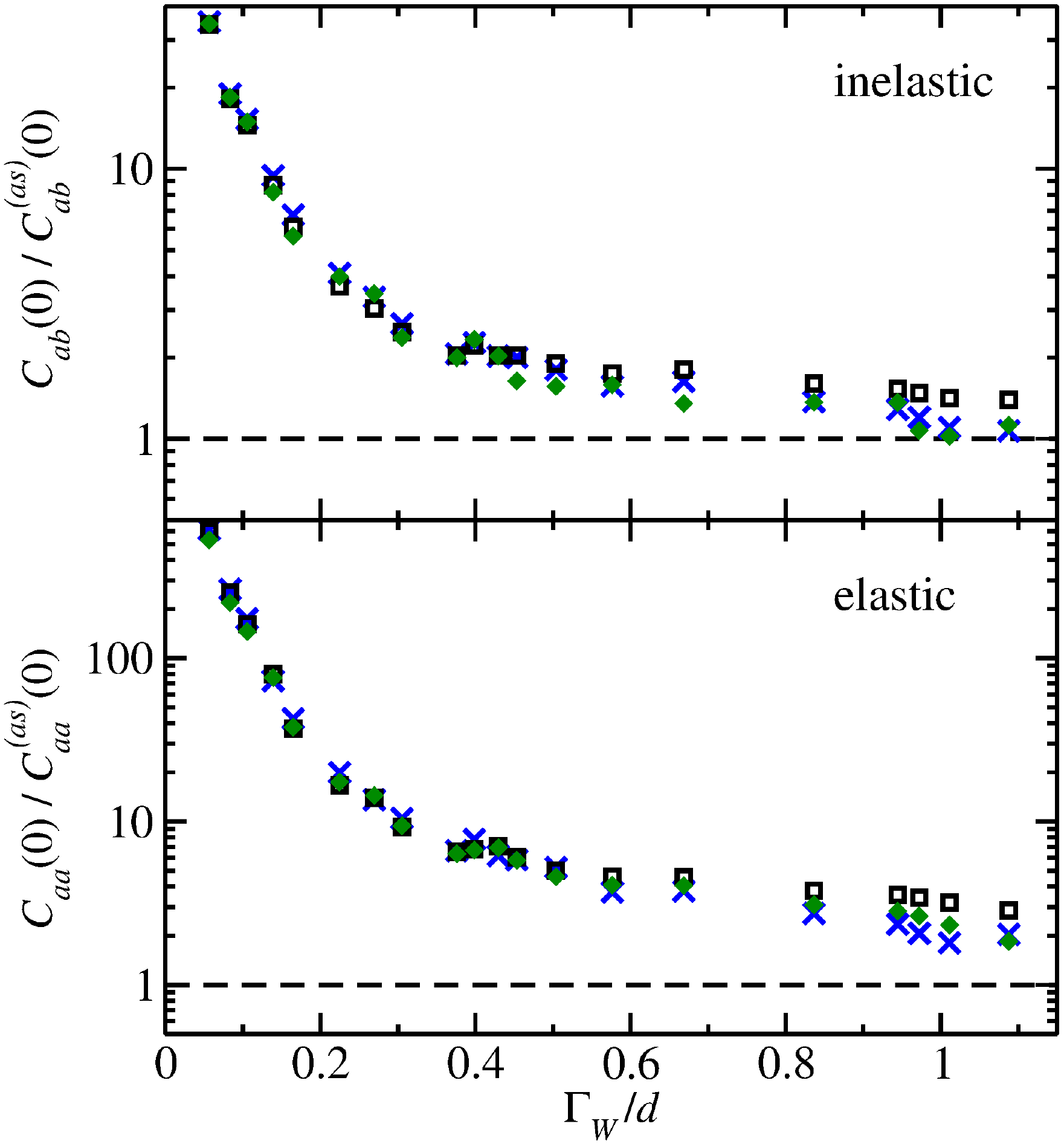}
        \caption{Ratios of $C_{ab}(\epsilon=0)$ to the corresponding $C^{(as)}_{ab}(\epsilon)$ defined in Eqs.~(\ref{Eq:Cabenergyas}) and~(\ref{Eq:Caaenergyas}), respectively. The transmission coefficients $T_1,T_2$ and $\tau_{abs}$ corresponding to the 20 values of $\Gamma_W/d$ considered in the figure are given in~\reftab{table1}. Black squares were obtained using the inverse Fourier transforms of~(\ref{Eq:CabtimeApp}) and~(\ref{Eq:CaatimeApp}) and of~(24) and~(25) in Ref.~\cite{ERI13}. Blue crosses show the exact analytical results given in Ref.~\cite{DAV88} and green diamonds the experimental ones. Upper panel: the inelastic case $a=1,\, b=2$. Lower panel: the elastic case for $a=b=1$. The scale has been chosen logarithmic for a better illustration of the good agreement between the different results.}
\label{fig5}
\end{figure}

\begin{figure}[ht!]
\includegraphics[width=0.9\linewidth]{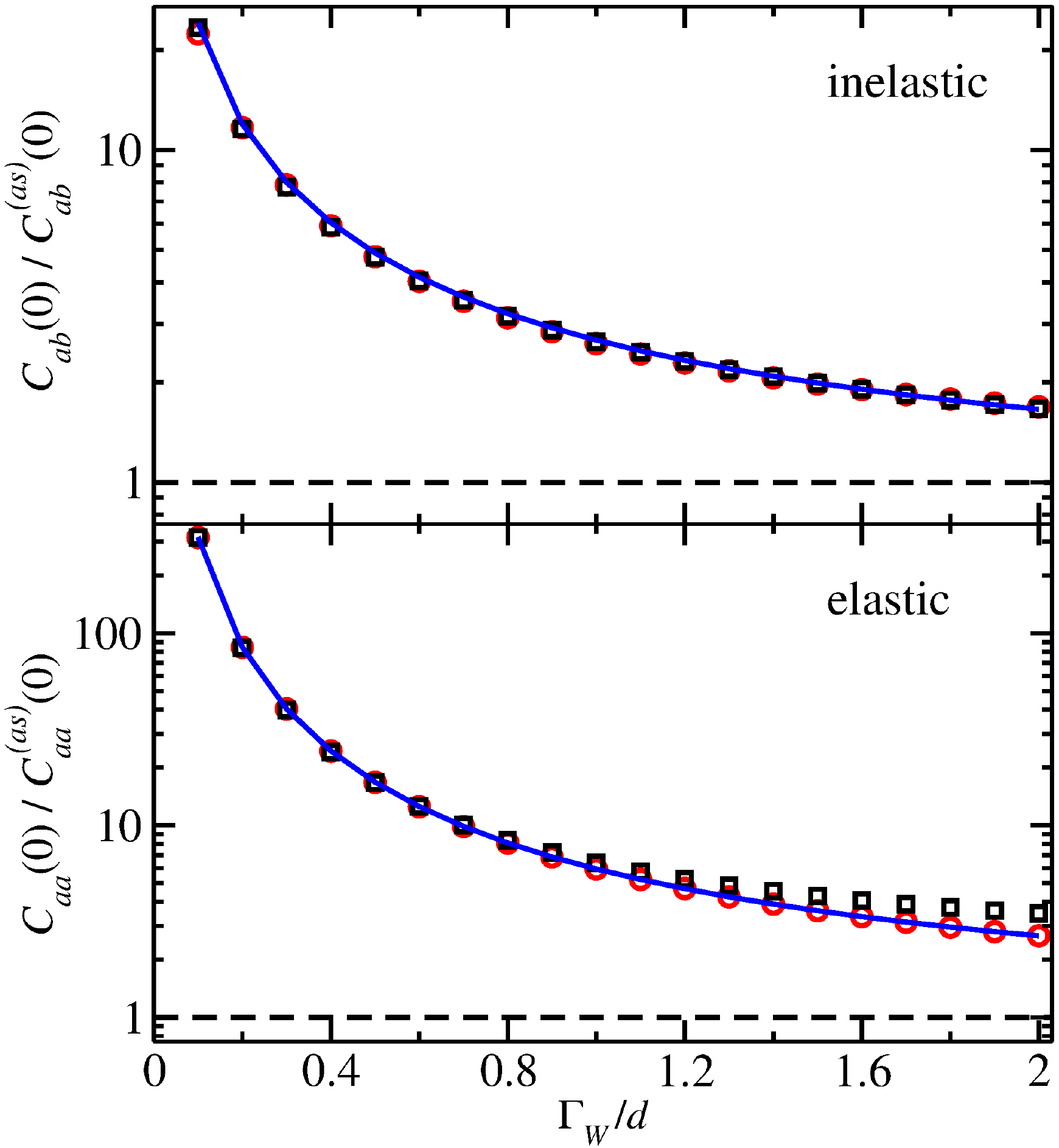}
\caption{
Ratios of $C_{ab}(\epsilon=0)$ to the functions $C^{(as)}_{ab}(\epsilon)$ defined in Eqs.~(\ref{Eq:Cabenergyas}) and~(\ref{Eq:Caaenergyas}), respectively, for 52 equal transmission coefficients. Black squares were obtained using the inverse Fourier transforms of~(\ref{Eq:CabtimeApp}) and~(\ref{Eq:CaatimeApp}) and of~(24) and~(25) in Ref.~\cite{ERI13}, red circles and the blue line show the RMT simulations and the exact analytical results given in Ref.~\cite{DAV88}, respectively. Upper panel: the inelastic case $a=1,\, b=2$. Lower panel: the elastic case for $a=b=1$. The scale has been chosen logarithmic for a better illustration of the good agreement between the different results, especially in the inelastic case.}
\label{fig2}
\end{figure}

The question of accuracy of the SBW model is an important issue in the following. It was tested by comparing results obtained with the model to experimental ones, to RMT-simulations and to the exact solution for the cross-sections at $\epsilon =0$~\cite{DAV88,DAV89,DIE10}. Figures~\ref{fig5}-\ref{fig2} show the ratios of the cross-section correlation coefficients~(\ref{Eq:fluctuationAApp}) and~(\ref{Eq:Cn=2elastic}) to $C_{ab}(\epsilon)/C_{ab}^{(as)}(\epsilon )$ at $\epsilon =0$ given in Eq.~(\ref{Eq:Cabenergyas}) and Eq.~(\ref{Eq:Caaenergyas}), respectively, as function of $\Gamma_W/d$. It is expected to approach the value unity, shown as dashed black line in the figures, for large values of $\Gamma_W/d$. We compare in Fig.~\ref{fig5} the SBW model (black squares) to RMT simulations (red circles) and experimental results plotted as green diamonds~\cite{DIE10}. For better visibility of the differences we show the ratios on a logarithmic scale. The inelastic case, shown in the upper panel of Fig.~\ref{fig5}, is relatively simple, since its only non-vanishing contribution to the decomposition (\ref{Eq:4pointcorrgeneralcrossA}) is the term  $C^{(4)}_{ab}(\epsilon)$ of Eq.~(\ref{Eq:C^4(epsilon)}).  In the SBW model it is defined by the three irreducible functions in Eq.~(\ref{Eq:fluctuationAApp}) given explicitly in Eqs.(\ref{Eq:2pointcorababenergy}),~(\ref{Eq:F_4App}) and~(\ref{Eq:G_4App}). For $\epsilon =0$, $C^{(4)}_{a\neq  b}(0)$ is the variance of the cross sections. For $\Gamma _W/d>1$ this variance becomes asymptotically the Ericson fluctuation term $| C ^{(2)}_{a\neq b} (0)|^2$, while as illustrated in Fig.~\ref{fig4}, it becomes the self-correlation term in the limit $\Gamma _W/d \rightarrow 0$. Deviations of the analytical model from the RMT simulations and the experimental results are largest for $\Gamma_W/d\gtrsim 0.8$. 

The elastic case is in much analogous to the inelastic one, but there are characteristic modifications most clearly seen in the SBW model. These appear since the average of any individual resonance to the $S$-matrix element no longer vanishes, even though the average partial width amplitude does so. The probability distribution of $S^{(fl)}(\epsilon)$ is no longer symmetric about the origin, which produces the three additional contributions, $\mathcal{H}_4^{aaaa}, \mathcal{F}_3^{aaa}$ and the two-point term $C^{(2)}_{aa}$; see Eq.~(\ref {Eq:Cn=2elastic}). Also in this case, shown in the lower panel of Fig.~\ref{fig5}, the SBW model closely reproduces the experimental and the RMT result but for a small systematic deviation for $\Gamma_W/d\gtrsim 0.8$. When reaching the experimentally achieved maximal value $\Gamma_W/d=1.09$, the ratios approach the values $1.5$ and $1.8$ for the inelastic case and the elastic one, respectively. Thus, there the cross-section correlation functions are already close to the correponding function $C^{(as)}_{ab}(\epsilon)$. Furthermore, the transition to the asymptotic value is slower in the latter case than in the former one, as already has been observed in Ref.~\cite{DIE10}. 

In order to allow a test of the SBW model for $\Gamma_W/d\gtrsim 1$ we performed additional RMT simulations and also evaluated the exact analytic result for the ratios~\cite{DAV88,DAV89,DIE10} up to $\Gamma_W/d = 2$. Here, we chose all 52 transmission coefficients equal. The results are presented in Fig.~\ref{fig2}. The deviations between the three different models seem to be very small even in the elastic case. They are visible only in the logarithmic scale used in the figure. Here, the agreement of the SBW model with the exact analytical results of Refs.~\cite{DAV88,DAV89} is better than that with the RMT simulations. In fact, it has been shown in Ref.~\cite{DIE10} that the agreement between the results of Refs.~\cite{DAV88,DAV89} and RMT simulations improves when choosing larger matrix dimensions in the $S$-matrix model~(\ref{eqn:Sab}) than those used here. Once more, the asymptotic values are not yet reached. As may be deduced from Figs.~\ref{fig4} and~\ref{fig4a}, and as has been found in Ref.~\cite{DIE10}, this limit is reached above $\Gamma_W/d\gtrsim\pi$. In view of the overall good agreement of the SBW model with experimental, exact analytical and RMT results we may conclude that it provides a good description for the variances of the cross-sections.
\subsection{Cross-section Autocorrelation Functions\label{Sec:Cross-sectioncorrfunction}}
\begin{figure}[ht!]
        \includegraphics[width=\linewidth]{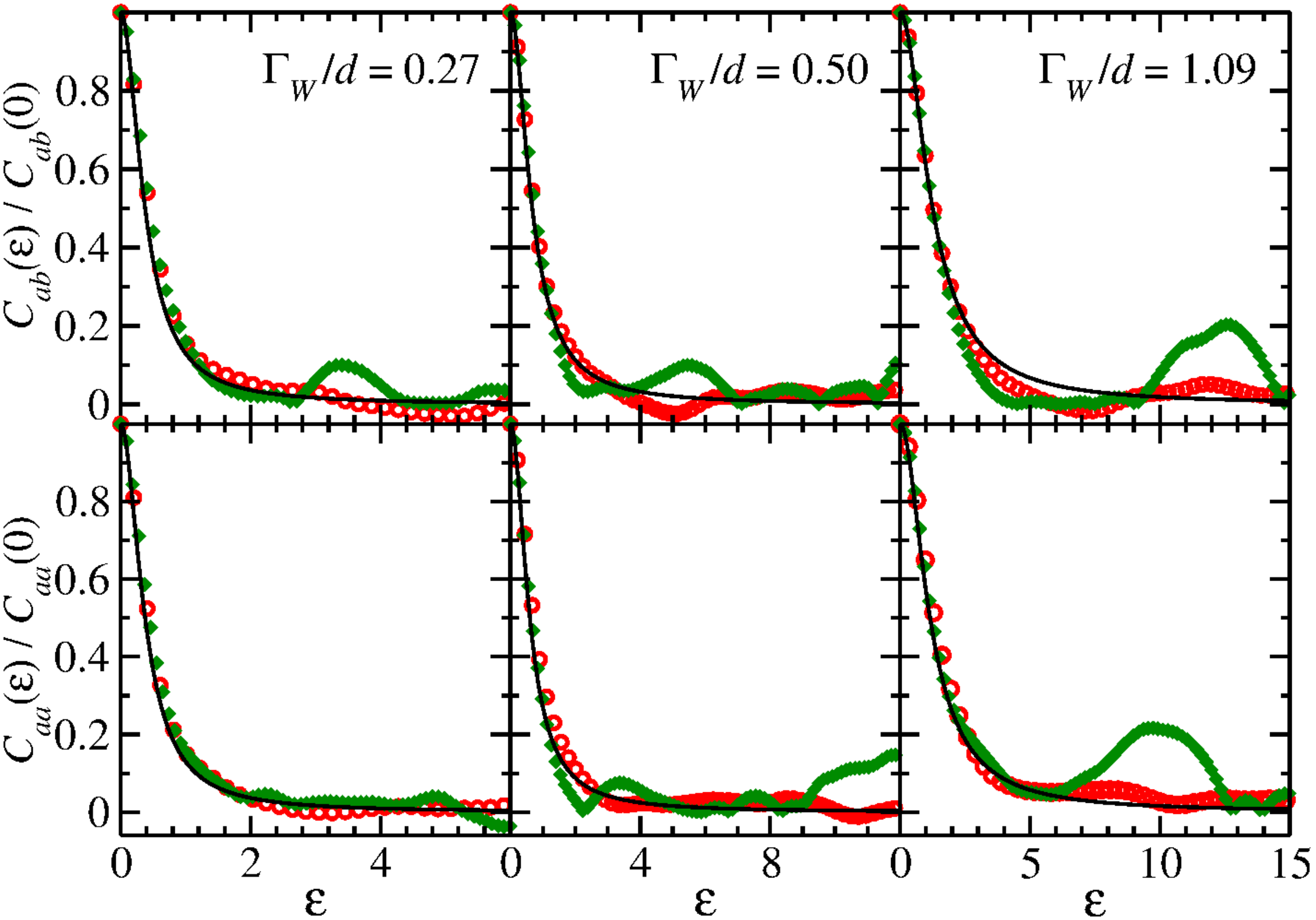}
        \caption{Cross-section correlation functions $C_{ab}(\epsilon)/C_{ab}(0)$. The 52 transmission coefficients $T_1,T_2$ and $\tau_{abs}$ corresponding to the values of $\Gamma_W/d$ in the insets are given in~\reftab{table1}. The full black lines were obtained using the inverse Fourier transforms of~(\ref{Eq:CabtimeApp}) and~(\ref{Eq:CaatimeApp}) and of~(24) and~(25) in Ref.~\cite{ERI13}. The red circles correspond to the RMT results, the experimental ones are plotted as green diamonds. Upper panel: the inelastic case $a=1,\, b=2$. Lower panel: the elastic case $a=b=1$.}
\label{fig6}
\end{figure}

Additional evidence for this last feature is given by Fig.~\ref{fig6}, which displays  $C_{ab}(\epsilon)/ C_{ab}(0)$ as a function of the energy increment $\epsilon $ for different values of $\Gamma_W/d$. The analytical results (black full lines) were obtained by computing the inverse Fourier transform of $\tilde C_{ab}(\tau)$ in Eqs.~(\ref{Eq:CabtimeApp}) and~(\ref{Eq:CaatimeApp}). For this we evaluated the integrals~(\ref{Eq:F_4timeApp})-(\ref{Eq:F_3timeApp}) and used Eqs.~(24) and~(25) from Ref.~\cite{ERI13} in order to determine the $S$-matrix autocorrelation functions. They are compared to RMT simulations (red circles) and experimental results (green diamonds). The agreement between SBW and RMT is striking. The bumps appearing in the experimental curves for $\epsilon\gtrsim 3-5$ are attributed to finite size effects. Note that they also appear in RMT simulations at an energy increment $\epsilon$ which depends on the size of the random matrices. Furthermore, the applicability of RMT to describe generic features in the long-range correlations of a system is justified only for $\epsilon$ values bounded by the length of the shortest periodic orbit. 
\begin{figure}[ht!]
        \includegraphics[width=\linewidth]{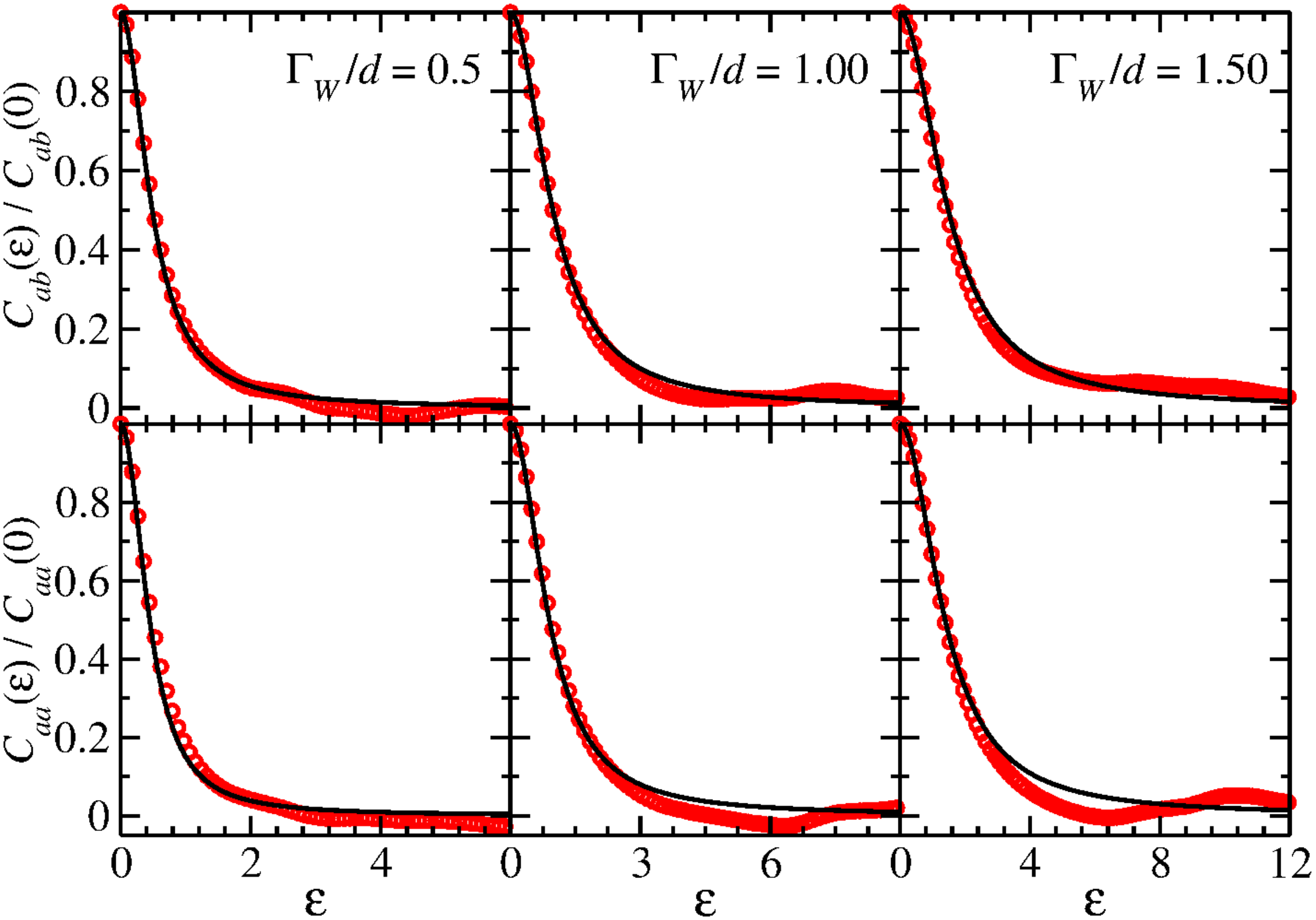}
        \caption{Cross-section correlation functions $C_{ab}(\epsilon)/C_{ab}(0)$ for $52$ equal transmission coefficients. The corresponding values of $\Gamma_W/d$ are given in the insets. The black full lines were obtained using the inverse Fourier transforms of~(\ref{Eq:CabtimeApp}) and~(\ref{Eq:CaatimeApp}), and of~(24) and~(25) in Ref.~\cite{ERI13}. Red circles illustrate the RMT results. Upper panel: the inelastic case $a=1,\, b=2$. Lower panel: the elastic case $a=b=1$.}
\label{fig3}
\end{figure}
Figure~\ref{fig3} shows a comparison of SBW results with RMT simulations. Once more, in the inelastic case the agreement between both models is very good. The widths of the SBW curves slightly underestimate that of the RMT simulations for $\Gamma_W/d\gtrsim 1$. The variations of the elastic and the inelastic cross-section autocorrelation functions with $\epsilon$ are nearly the same. However, the widths of the former ones are smaller than those of the latter ones.   
%
%SSSSSSSSSSSSSSSSSSSSSSSSSSSSSSSSSSSSSSSSSS
   \subsection{The Self-Correlation Terms\label{subsection:widths}}
   %%%%%%%%%%%%%%%%%%%%%%%%%%%%%%%%%%%%%%%%%%
   The results for the inelastic and elastic cases demonstrate that the self-correlation term (\ref{Eq:F_4App}) is dominant for a small total width $\langle\Gamma\rangle /d$, i.e., Weisskopf width $\Gamma_W/d$, and provides the major contribution to the cross-section correlations for $\Gamma_W/d\lesssim\pi^{-1}$. In that region, the cross-section coefficients accurately reproduce the exact results; see Fig.~\ref{fig2}. The features of the self correlations become apparent in the SBW model, which provides explicitly the dominant contribution responsible for the underlying mechanism. The self correlations are accounted for by the first term in Eq.~(\ref{Eq:F_4App}), 
\begin{equation}
\label{Eq:selfcorr4ab}
\Xi^{(4)}_{ab}=4\pi\left\langle\frac{\Gamma _{a}^2\Gamma _{b}^2}{\Gamma^3}\right\rangle . 
\end{equation}
Assuming a Porter-Thomas distribution for the partial widths and their ratios $x_e=\Gamma_e/\langle\Gamma_e\rangle$, it can be computed explicitly. It takes the simplest form for the case of equal transmission coefficients, where it is given for the elastic case $a=b$ in terms of the ratio $\rho=\langle\Gamma_a\rangle/\Gamma_W$ by (see also Ref.~\cite{GOR02})
\begin{equation}
\label{Eq:selfcorr4aa}
\Xi^{(4)}_{aa}\simeq 4\pi\frac{105\rho^4}{(1+6\rho)(1+4\rho)(1+2\rho)}\Gamma_W/d.
\end{equation}
For a constant total width $\Gamma_W/d$, this self-correlation term varies rapidly with the 4th power of the average partial width $\langle\Gamma_a\rangle$, whereas for a constant ratio $\rho$ it changes only linearly with $\Gamma_W/d$. Note, that for equal transmission coefficients the ratio $\rho\equiv 1/\Lambda$ is constant. In the microwave experiments the partial widths of the fictitious channels were equal and differed from those of the antenna channels, which both take similar values and the ratios $\langle\Gamma_{1,2}\rangle/\Gamma_W$ were also approximately constant in the frequency intervals 10-19 GHz and 20-25 GHz, respectively. The corresponding values are given in the 6th column of \reftab{table1}. 

\begin{figure}[ht!]
        \includegraphics[width=0.9\linewidth]{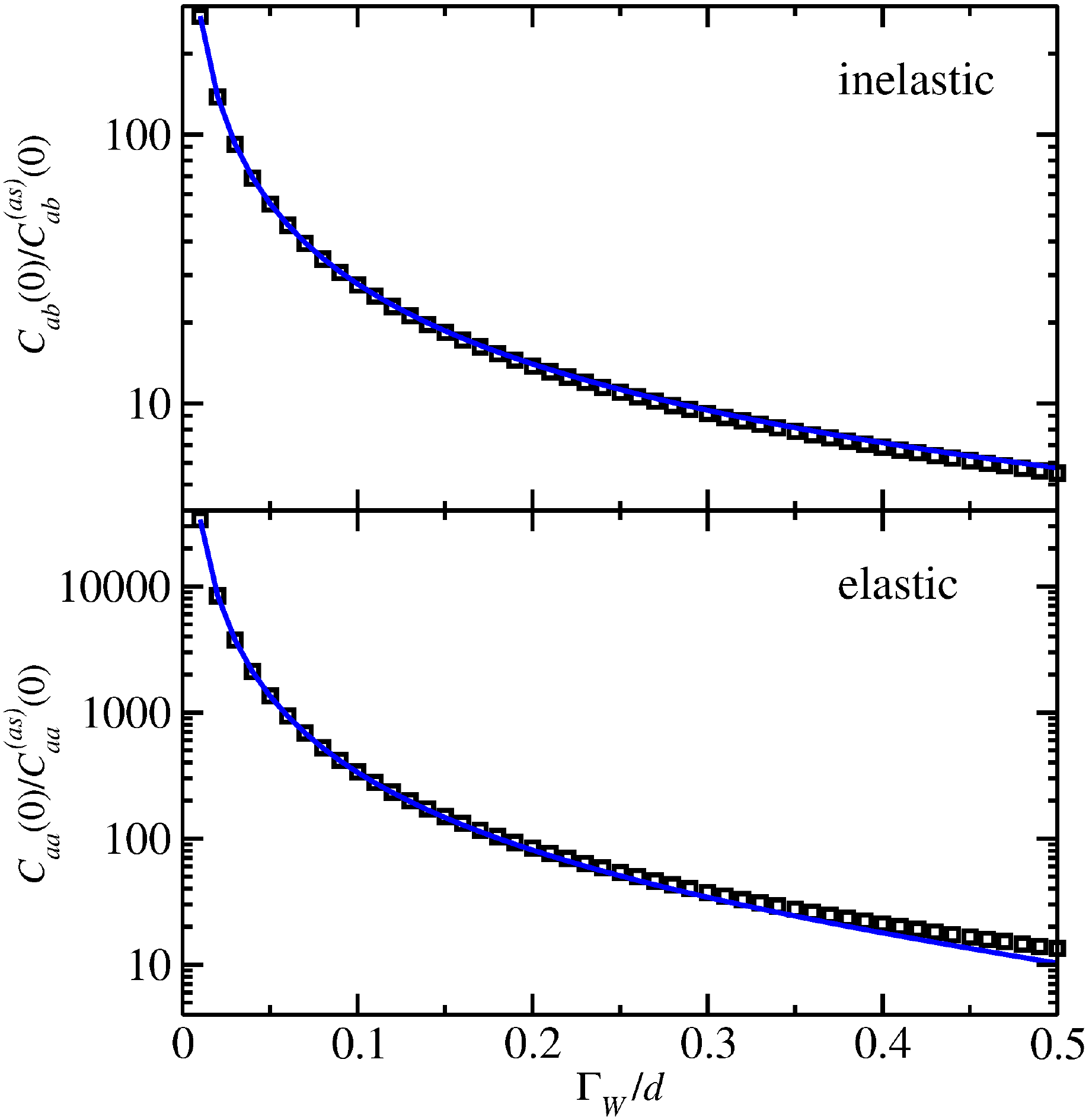}
        \caption{
Ratios of $C_{ab}(\epsilon=0)$ to $C_{ab}^{(as)}(\epsilon=0)$ defined Eqs.~(\ref{Eq:Cabenergyas}) and~(\ref{Eq:Caaenergyas}). Here, 202 open channels with equal transmission coefficients were taken, corresponding to a small $\rho =1/202$. Red squares show the approximations~(\ref{ratioapproxaa}) and~(\ref{ratioapproxab}) obtained by taking into account only the self correlations for the elastic case (lower panel) and the inelastic case (upper panel), respectively. The blue line shows the exact analytical results obtained from Ref.~\cite{DAV88}. A logarithmic scale was chosen to better illustrate their good agreement up to $\Gamma_W/d\simeq 0.2$.}
\label{fig8}
\end{figure}

The lower panels of Figs.~\ref{fig5} and~\ref{fig2} show the ratio $C_{aa}(0)/C^{(as)}_{aa}(0)$. The asymptotic cross-section correlation function $C^{(as)}_{aa}(0)$ is given in~(\ref{Eq:Caaenergyas}). The self-correlation term associated with $C^{(2)}_{aa}(0)$ corresponds to the first term in Eq.~(\ref{Eq:2-pointaabb}). Proceeding as in Eq.~(\ref{Eq:selfcorr4aa}), it can also be computed explicitely, yielding  
\begin{equation}
\label{Eq:selfcorr2aa}
\Xi^{(2)}_{aa}=2\pi\left\langle\frac{\Gamma_{a}^2}{\Gamma}\right\rangle =2\pi\frac{3\rho^2}{(1+2\rho)}\Gamma_W/d. 
\end{equation}
Using the SBW results Eqs.~(\ref{Eq:selfcorr4aa}) and~(\ref{Eq:selfcorr2aa}) yields with $C_{aa}^{(as)}(0)\simeq 2\pi^2\rho^2(\Gamma_W/d)^2C^{(2)}_{aa}(0)$ for small $\rho$ and $\Gamma_W/d$ 
\begin{equation}\label{ratioapproxaa}
\frac{C_{aa}(0)}{C_{aa}^{(as)}(0)}\simeq\frac{1}{3\pi^2}\frac{105}{(1+4\rho)(1+6\rho)}\frac{1}{(\Gamma_W/d)^2}.
\end{equation}
Similarly, we obtained for the inelastic self-correlation terms, assuming equal transmission coefficients, which implies $\langle\Gamma_a\rangle =\langle\Gamma_b\rangle$ for the observed channels $a,b\in\{1,2\}$,
\begin{equation}\label{Eq:selfcorrab}
\Xi^{(4)}_{ab}=36\pi\frac{\rho^4}{(1+6\rho)(1+4\rho)(1+2\rho)}\Gamma_W/d,
\end{equation}
\begin{equation}
\label{Eq:selfcorrab2}
\Xi^{(2)}_{ab}=2\pi\frac{\rho^2}{(1+2\rho)}\Gamma_W/d
\end{equation}
and, accordingly, for the inelastic ratio
\begin{equation}\label{ratioapproxab}
\frac{C_{ab}(0)}{C_{ab}^{(as)}(0)}\simeq\frac{9}{\pi}\frac{1+2\rho}{(1+6\rho)(1+4\rho)}\frac{1}{\Gamma_W/d}.
\end{equation}
For large numbers of open channels $\Lambda$, i.e., small values of $\rho$, the approximate results in Eqs.~(\ref{ratioapproxaa}) and~(\ref{ratioapproxab}), obtained by considering only the self-correlation terms~(\ref{Eq:selfcorr4aa}),~(\ref{Eq:selfcorr2aa}),~(\ref{Eq:selfcorrab}) and~(\ref{Eq:selfcorrab2}) depend moderately on $\rho$. In Fig.~\ref{fig8}, we compare them with the exact analytical results obtained from Ref.~\cite{DAV88}. The agreement is good, both in the elastic and the inelastic case, for $\Gamma_W/d\lesssim 0.2$. 

The sensitivity of the self-correlation terms with respect to the values of $\langle\Gamma_a\rangle/d$ and above all the close agreement of the SBW results with exact ones for this range of small $\langle\Gamma_a\rangle/d$ and narrow resonance widths provide a strong indirect evidence for the persistence of the normal width distribution in the tail of the corresponding distribution. This conclusion extends to the experimental situation as well. There again the close agreement with exact calculations in the region of dominant self correlations emphasizes the validity of a normal width distribution. This sensitivity is of practical importance since it implies that experiments must ensure that the data samples are sufficiently large so as not to distort the results. Note, that in the region of isolated resonances the contribution of the average partial widths of the observed channels, $\langle\Gamma_a\rangle$ with $a=1,2$, to the total width $\Gamma_W$ are approximately $6-10\%$ (see~\reftab{table1}). Moreover, only a small fraction of less than $1\%$ of the total number of states contribute to the self-correlations. Consequently, in this region the correlations are produced by widely spaced resonances with an exceptionally strong partial width in the observed channels. These observations will be further elaborated in a future publication. 
  
   %SSSSSSSSSSSSSSSSSSSSSSSSSSSSSSSSSSSSSSSSSSSSSSSSSSSSSSSSSSSSSSSSS
\section{Conclusion}
The aim of  the present article has been to obtain an analytical approximation to the cross-section correlations and related functions for a chaotic scattering system. The analysis was based on the SBW model, extensively used previously in the asymptotic limits of narrow and strongly overlapping resonances. Here, we have set three goals (i) firstly, to corroborate that the SBW model provides quantitative results for the $S$-matrix autocorrelation functions close to exact ones under rather general conditions (ii) secondly, to establish the general analytical expressions for the four-point $S$-matrix correlation functions (iii) thirdly, to demonstrate that these results give new insights into the physical mechanisms  which produce the correlations as well as to illustrate the sensitivity of the correlation functions to their input parameters.  As will be discussed below, the goals set in the introduction have been achieved and the results exceed our initial expectations.\\
 
(i) The accuracy of the SBW model is already indicated by its similarity to an approximation derived analytically based on the exact results~\cite{VER85} in Ref.~\cite{ERI13} which was shown to provide a good description for the $S$-matrix correlation functions. Figure~\ref{fig1} illustrates this agreement between the approximations and the exact analytical results for different experimentally relevant situations. It emphasizes the importance of the observed channels for the overall scale and the insensitivity to the details of the remaining ones. The unitarity constraint on the $S$-matrix results in the optical theorem which is accounted for on the average. Once this is the case, the lack of unitarity of the $S$-matrix approach Eq.~(\ref{Eq:S-abres}) is of little importance for observed channels $a,b$ with $\langle\Gamma_{a,b}\rangle /\Gamma_W\ll 1$ since in the overlap region up to this correction the variances of the $S$-matrix correlations are identical for the SBW model and the VWZ approach, which preserves unitarity. Moreover, in the overlap region the cross-section variances are well described by $S$-matrix variances as illustrated in Figs.~\ref{fig4} and~\ref{fig4a}. We therefore conjecture that the unitarity corrections are equally small in this case. The good agreement of the experimental, numerical and analytical results compared in Figs.~\ref{fig5} and \ref{fig2} further emphasizes empirically that the SBW model captures the essence of the exact results correctly also for weakly overlapping and isolated resonances.\\
 
(ii) The second goal of deriving closed analytical expressions for the cross-section correlation functions in the SBW model was considered  unachievable in Ref.~\cite {GOR02}. Our results are obtained by first observing that they are special cases of the four-point $S$-matrix correlation functions. These are more conveniently expressed in terms of irreducible four-point, three-point and two-point correlation functions, yielding a modification of the expansion given in Ref.~\cite{DIE10}. In general there are three such irreducible four-point functions, two once reducible ones and one which is twice reducible. This expansion is particularly complex for correlations between elastic cross sections. The different irreducible terms are further simplified by appropriate contractions taking into account the level cluster correlations which depend on the overall conditions. Here, we restrict the discussion to the standard Dyson level correlations~\cite{DYS63}. Terms depending on the three- and four-level cluster functions have been neglected since they correspond to a simultaneous close encounter of more than two levels which is intuitively unlikely to occur. Results are given in~\refsec{Sec:irreducible} in the energy and the time representations. They are formulated very generally and cover all cases from a single channel to a large number of open channels with a large variety of transmission coefficients from the region of narrow resonances to the asymptotic region of overlapping ones. 

The accuracy of the analytical SBW results for the cross-section correlation functions has been tested by comparison to exact results for the cross-section variances in Figs.~\ref{fig5} and~\ref{fig2} and to experimental and RMT results for the cross-section correlation functions in Figs.~\ref{fig6} and~\ref{fig3}. The deviation observed in Fig.~\ref{fig5} for the smallest value of $\Gamma_W/d$ may be explained by the small number of resonances in the corresponding 1~GHz window; see~\reftab{table1}. We conclude from the agreement between the SBW model calculations and the experimental and RMT results that below $\Gamma_W/d\lesssim 2$ they are nearly independent of the unitarity constraint. This is not surprising because it is largely imposed via the normal Breit-Wigner form of an individual resonance with an average background amplitude. As observed above, the model is conjectured to closely reproduce exact results in the region of large $\Gamma_W/d$. Note that in Refs.~\cite{AGA75,WEI84,DIE10} analytical expressions were derived for the cross-section correlation functions that are applicable beyond the value $\Gamma_W/d\sim 2$, where the distributions of the real and imaginary parts of the $S$-matrix elements have Gaussian distributions~\cite{Fyodorov2005,Kumar2013}.

Like in the VWZ model the present approach tacitly assumes that during the scattering process a long-lived (quasibound) chaotic state has been produced in a short time compared to its life-time. Since the life-time of the chaotic state is inversely proportional to the Weisskopf width this automatically implies that it has a maximum value set by the formation time. The latter depends on the dynamical properties in the interaction region. For an initially closed system with chaotic dynamics it can be identified with the length of the shortest periodic orbit~\cite{Bluemel1992,Jung1997}. In the case of an intermediate motion in an optical potential as in nuclear physics it can be deduced from the widths of the corresponding resonances and their decay to doorway states.

An interesting by-product is that they apply equally well to any distribution of partial width amplitudes with random sign. We have not explored their sensitivity to the explicit form of this probablity distribution, which has always been assumed to be the normal one. We note, that the SBW model is in this respect more flexible than the RMT models which assume randomness on the level of the initial interaction with a normal distribution of the matrix elements of the associated Hamiltonian. Consequently, the second goal even exceeded our expectations concerning the capability of the SBW model. 
\\

(iii) Our third and principal goal was to improve the understanding of the different contributions generating the correlation functions and of their relevance, because the RMT approach gives only  global results. The SBW results are very general covering the whole range of isolated to overlapping resonances. In view of the large variety of situations we focus on the important special case for which none of the partial widths that constitute the Weisskopf width dominates. The latter approximately yields the total width. 
 
 We consider two examples. The first one concerns the consequences of the transition from the region of non-overlapping, narrow resonances to the asymptotic region of overlapping ones. Here the simplest situation is the inelastic cross-section autocorrelation function illustrated in the upper panel of Fig.~\ref{fig4}, which demonstrates the change from dominance of self-correlations for $\Gamma _W/d\lesssim\pi^{-1}$ and even beyond to that of the Ericson fluctuation term with increasing $\Gamma _W /d$. The sum of these two contributions gives an excellent description in the entire range of $\Gamma _W/d$. 
  
The corresponding elastic situation is illustrated in the lower panel of Fig.~\ref{fig4}. As for the inelastic case the self-correlations dominate for $\Gamma _W/d\lesssim\pi^{-1}$ whereas contributions from the remaining four- and three-point correlations cancel each other systematically and substantially. The region between $1<\Gamma _W<2 $ is characterized by comparable contributions from self-correlations and from the asymptotic two-point correlation term, while for larger $\Gamma _W$ the latter rapidly becomes dominant.  The Ericson fluctuation term is everywhere small. The conclusion to be drawn from these observations is thus that for $0.5\lesssim\Gamma _W/d\lesssim 4$ the sum of the self-correlations and the asymptotic two-point term approximate the cross-section autocorrelations well in the elastic case and for larger values of $\Gamma _W/d$ only the two-point correlations survive. 

We also computed the cross-section correlation functions as function of $\epsilon$ and compared them to RMT results obtained via numerical simulations. Figures~\ref{fig6} and~\ref{fig3} demonstrate that the latter, normalized to unity at $\epsilon =0$, are accurately described by the SBW model and are insensitive to its details. We conclude from these observations that the central features governing chaotic cross-section correlations -- as well as $S$-matrix correlations -- are determined by the distributions of the partial width amplitudes of the observed channels. 
  \\
The SBW model provides explicitly the functional dependence of the correlation functions on the level correlations. For narrow and weakly overlapping resonances the contributions to the elastic cross-section variances are typically dominated by self-correlations. They are mainly produced by widely spaced states with exceptionally large width in the observed channels. This feature is well supported by the agreement of the SBW model results with the corresponding exact ones, illustrated in Figs.~\ref{fig2}-\ref{fig8}. The agreement with experimental data demonstrates that the Porter-Thomas width distribution indeed is valid over a large range for the underlying probability distribution in Eq.~(\ref{Eq:selfcorr4ab}) and implied experimentally as well by Fig.~\ref{fig5}.

We note that the assumption that no partial width dominates can be achieved in two physically distinct ways. Firstly, it can be realized in systems where the non-observed channels are either due to incoherent absorption or to a large number of channels sufficiently weak so as not to produce noticeable mixing effects, secondly by means of many channels with transmission coefficients close to unity and very large Weisskopf widths as discussed in Refs.~\cite {SOK89,CEL07,CEL08,SOR09,DIC54}. In the latter case these can coherently produce narrow ("trapped") states as well as broad ("super-radiant") ones. Such phenomena are of different nature and beyond the framework of the present article.\\
  
  Finally, we note that the approach considered in the present article can be generalized in a number of ways. In particular, we provide explicit expressions for the cross-section correlation functions of systems with width distributions different from the Porter-Thomas one, although we have not explored the sensitivity of the results with respect to their choice. Another generalization would be the extension to systems with a broken symmetry or violated time-reversal invariance. 
  
  %  \\   %{
\begin{acknowledgments}
This work was supported by the Deutsche Forschungsgemeinschaft (DFG) within the Collaborative Research Center 634. BD and AR are grateful for the hospitality during their stay at CERN. We thank H. A. Weidenm\"uller for constructive comments. BD is grateful to T. Gorin and to T. Seligman for helpful discussions.
\end{acknowledgments}
\begin{widetext}
%%%%%AAAAAAAAAAAAAAAAAAAAAAAAAAAAAAAAAAAAAAAAAAAAAAAAAAAAAAAAA
\section{APPENDICES}
%AAAAAAAAAAAAAAAAAAAAAAAAAAAAAAAAAAAAAAAAAAAAAAAAAAAAAAAAA
\appendix       % {APPENDICES}
%AAAAAAAAAAAAAAAAAAAAAAAAAAAAAAAAAAAAAAAAAAAAAAAAAAAAAA
\section{Distributions and Notations\label{App:Distributions}}
%AAAAAAAAAAAAAAAAAAAAAAAAAAAAAAAAAAAAAAAAAAAAAAAAAAAAAAAAA
\subsection{General Probability Averages and Short-Hand Notation \label{App:shorthand}}
%AAAAAAAAAAAAAAAAAAAAAAAAAAAAAAAAAAAAAAAAAAAAAAAAAAAAAAAAAA
Consider a probability distribution $p(x)$ with the average  $\langle x\rangle =1$. Define the quantity  $A(k)\equiv\langle x^k\rangle $. Then
 \begin{eqnarray}
   \hspace{-0.3in}&&   
   \langle x^{k_e } \exp(-\lambda T_e  x) \rangle \equiv \langle x^{k_e} \rangle {\langle  x^{k_e} \exp(-\lambda
   T_e  x) \rangle \over \langle x^{k_e} \rangle }\equiv A(k_e)g_e (\lambda T_e). ~~~~~~~
\end{eqnarray}
     In the separable description define
\begin{eqnarray}
 \hspace{-0.3in}&& 
\Pi _{e; abc..}(\lambda )=\prod _e \langle x^{k_e} \exp(-  \lambda T_e  x) \rangle \equiv \prod _{e} A(k_e)\, g_e (\lambda T_e ). \label{Eq:Pi_eabc}
\end{eqnarray}
Here, the product is over all open channels $e$ with transmission coefficients $T_e$. The labels $l=a,b,c,\cdots$ correspond to those of the $S$-matrix elements in the correlation function under consideration. Each such label contributes one unit to the corresponding exponent $k_l$ of $x$, that is, $k_l$ equals the number of occurrences of an index $l$. The open channels $e$, that do not coincide with one of the labels $l$ have $k_e=0$ and $A(0)=1$. 

For the special case of the generalized Porter-Thomas distributions the probability distribution has the form 
\begin{eqnarray}
\hspace{-0.3in}&& \label{App:Genprobabilitydistr}
p_{\nu }(x)=\Gamma (\nu )^{-1}\nu ^{\nu }  x^{\nu -1}\exp (-\nu x)
\end{eqnarray} 
in terms of the Gamma-function $\Gamma (\nu )$ with $\langle x \rangle =1$  and $\langle x^2-\langle x \rangle ^2\rangle =1/\nu$. For such distributions Eq.~(\ref{Eq:Pi_eabc}) becomes
\begin{eqnarray}
\hspace{-0.3in}&&\label{Eq:Anuk}
\langle x^{k_e}\exp (-\lambda T_e x)  \rangle _{\nu }=A_{\nu}(k_e)\left(1+\lambda  T_e/\nu\right)^{-(k_e+\nu)}
\end{eqnarray} 
with $A_{\nu}(k)={\Gamma (\nu +k)\over \Gamma (\nu )\nu ^k}$ the generalized Porter-Thomas enhancement factor.  Special values  are: 
\begin{eqnarray}
&A_{1/2}(k)=(2k-1)!!~~~ &{\rm Porter-Thomas\, distribution}\nonumber\\ 
&A_1(k)=k!~~~ &{\rm exponential\, distribution}\nonumber\\ 
&A_{\nu=\infty }(k)=1~~~ &{\rm constant\, width}\nonumber
\end{eqnarray}
For a generalized Porter-Thomas distribution  Eq.~(\ref{Eq:Pi_eabc}) becomes 
  \begin{eqnarray}
  \hspace{-0.3in}&&\label{Eq:shorthandsumgeneral}
   \Pi_{e;ab..}^{(\nu )}(\lambda )\equiv  
    A_{\nu}(k_{a })\left(1+\lambda  T_{a }/\nu\right)^{-k_{a }
    }A_{\nu}(k_{b})\left(1+\lambda  T_{b}/\nu\right)^{-k_{b }
    }..\prod _e(1+\lambda T_e/\nu )^{-\nu }.~~~~~~~~~~~~~~~~~~~~~~~~~
   \end{eqnarray}
The standard Porter-Thomas distribution~\cite{POR56} corresponds to  $\nu =1/2$.
\subsection{ The Two-Level Cluster Function \label{App:spacingfunctions} }
The two-level cluster function is denoted by  $Y_2(r)$, with $r$ the spacing between adjacent energy levels. The levels are rescaled to average spacing unity. The form factor $b(\tau)$ corresponds to the Fourier transform of $Y_2(r)$, $b(\tau )=\int Y_{2}(s)\exp (2\pi i\tau  r)dr$~\cite{MEH67}. For the GOE it is given by the Dyson expression~\cite{DYS63}
\begin{eqnarray} 
 \hspace{-0.3in}&& 
  b(\tau )= 1-2|\tau |+|\tau||\ln [1+2|\tau|] \simeq 1-2|\tau |+2|\tau |^2-2|\tau | ^3+8|\tau |^4/ 3-4\tau |^5+\cdots ~0< |\tau |  <1,     \nonumber \\
 \hspace{-0.3in}&&  \label{Eq:Dyson}
   b(\tau )= -1+|\tau |\ln \left[{2|\tau |+1 \over |2\tau |-1}\right]\simeq 0+(1/ 12)|\tau |^{-2}+\cdots~~ ~|\tau| >1.
   \end{eqnarray} 
  Approximate expressions for these functions respecting scales and normalization are
    \begin{eqnarray} 
 \hspace{-0.3in}&& 
 Y_{2}(r)\simeq  (1+(\pi r)^2)^{-1}; ~~~~~b(\tau )\simeq \exp (-2|\tau |) . \label{Eq:clusterapprox}
  \end{eqnarray} 
  
 \section{Some General Results for Correlation Functions \label{App:Results}}
 The correlation functions are given in a general form as follows:
  \begin{eqnarray}
  \hspace{-0.3in}&& 
  C[f,g](\omega)\equiv  \label{Eq:Defcorrfunc}
 \left \langle f(x_0-{\omega \over 2})g(x_0+{\omega \over 2})\right \rangle  -\left \langle f(x_0)\right \rangle \left \langle g(x_0)\right \rangle  
 \end{eqnarray} 

 \subsection {Two-Point $S$-Matrix Correlations \label{Sec:2pointgen}}
The full two-point function in the energy representation equals for the inelastic case $a\ne b$
\begin{equation}
\hspace{-0.3in}
C^{(2)}_{ab}(\epsilon) =  \label{Eq:2pointfullenergyab}
2\pi \left \langle {\Gamma _{a}\Gamma _{b}\over i\epsilon +\Gamma } \right \rangle ,  
\end{equation}
and for the elastic one $a=b$
\begin{equation}
\hspace{-0.3in}
C^{(2)}_{a a}(\epsilon) =  \label{Eq:2pointfullenergyaa}
  2\pi ~\left[ \left \langle {\Gamma _{a}\Gamma _{b}\over i\epsilon +\Gamma }\right \rangle -
 \int_{-\infty }^{\infty }{\rm d}r Y_2(r) \left \langle {\Gamma _{1a}\Gamma _{2b}\over i(\epsilon -r)+(\Gamma _1+\Gamma _2)/2} \right \rangle  \right].
\end{equation}
The Fourier transforms for the inelastic case~(\ref{Eq:2pointfullenergyab}) is given as 
\begin{equation}
\hspace{-0.3in}   \label{Eq:2pointfulltimegeneralinel}
\tilde{C}^{(2)}_{ab}(\tau )=(2\pi )^2 
\left \langle \Gamma _{a}\Gamma _{b  }\exp \left(-2\pi  \tau \Gamma \right)\right \rangle , 
\end{equation}
and for the elastic one~(\ref{Eq:2pointfullenergyaa}) as
\begin{equation}
\hspace{-0.3in}  
\tilde  {C}[S_{a a}S_{bb}^*](\tau )=(2\pi )^2\label{Eq:2pointfulltimegeneralel}
 \left( \left \langle \Gamma _{a}\Gamma _{b  }\exp \left(-2\pi \tau \Gamma \right)\right \rangle   -b(\tau )\left \langle \Gamma _{1a}\Gamma _{2b}\exp(-2\pi\tau (\Gamma _1+\Gamma _2)/2 )\right \rangle \right).
  \end{equation}
In both cases it is non-vanishing for $\tau >0$.
     
In the separable short hand notation of Appendix~\ref{App:shorthand} this gives for the inelastic case       
\begin{equation}
\hspace{-0.3in}    
\tilde{C}^{(2)}_{ab}(\tau)= T_aT_b \Pi _{e;ab}(\tau ),
\label{Eq:2pointfulltimeishortinel}
\end{equation}
and for the elastic one
\begin{equation}
\hspace{-0.3in}   \label{Eq:2pointfulltimeshortel}  
 \tilde{C}[S_{a a}S^*_{bb}](\tau)=  T_aT_b \Big( \Pi _{e;ab}(\tau )  -   b(\tau) \Pi _{e;a}(\tau /2)\Pi _{e;b}(\tau /2)\Big).
\end{equation}
The exponential approximation to the time-variation is valid as long as $(2\pi\tau/d)^2 \left\langle (\Gamma - \langle \Gamma  \rangle )^2\right\rangle<1$.  In terms of the generalized Porter-Thomas distributions given in Appendix \ref{App:Distributions} this condition becomes $\tau^2\sum _eT_e^2/(2\nu) <1$. For the elastic case the variation is additionally modulated by the form factor $b(\tau ) $. The deviation converges to zero for increasing $\nu\to\infty$. The distribution becomes an exponential one in this limit of a constant transmission coefficient.

\subsection{Time Correlation Functions \label{Sec:Timecorrelations} }
The time correlation functions $\tilde {C} [\sigma _{ab}\sigma _{cd}](\tau ) $ are the Fourier transforms of the corresponding ones in the energy representation; see Eqs.~(\ref{Eq:4pointcorrfunction}) -~(\ref{Eq:F_3App}). We consider only autocorrelation functions so that the indices $c,d$ take the values $a~ {\rm or} ~b$ only. The corresponding expressions in the short-hand notation are given in Eqs.~(\ref{Eq:F_4timeApp}) -~(\ref{Eq:F_3timeApp}).

The Fourier transform  of $\mathcal{F}^{abcd} _4(\epsilon )$ of Eq.~(\ref{Eq:F_4App}) is
\begin{eqnarray}
\hspace{-0.3in}&&
\tilde{\mathcal{F}}_4^{abcd}(\tau )= \label{Eq:F_4irrtimeAApp} 
\int ^{\infty}_{-\infty} d\epsilon \exp(2\pi i \epsilon \tau ) \mathcal {F}_4^{abcd}(\epsilon) = (2\pi )^2  
\Big[\left \langle { \Gamma _{1a}\Gamma _{1b}\Gamma _{1c}\Gamma _{1d} \over \Gamma _{1}^2}\exp(-2\pi \Gamma _{1} |\tau | )\right \rangle  -~~~~
\\
\hspace{-0.3in}&& \nonumber 
 ~b(\tau )  ~
 \left \langle { \Gamma _{1a}\Gamma _{1b}\over \Gamma _1}exp(-\pi \Gamma _1|\tau |  )\right  \rangle \left \langle {\Gamma _{2c}\Gamma _{2d} \over \Gamma _2}\exp(-\pi \Gamma _2|\tau |  )\right \rangle \Big].
\end{eqnarray}
The Fourier transform  of $\mathcal{G}^{abcd}_4(\epsilon )$ of Eq.~(\ref{Eq:G_4App}) has at most two differing indices (a,b) corresponding to the cases (ab;ab) and (aa;bb)\\
\begin{eqnarray}
\hspace{-0.3in}&& 
 \tilde{\mathcal{G}}_{4}^{abcd}(\tau  )=  
\Big \{  (2\pi )^3 \left \langle  {\Gamma _{1a}\Gamma _{1b}\Gamma _{2a}\Gamma _{2b}
 \over \Gamma _1+\Gamma _2 } \Big(\exp(-2\pi \Gamma _1 |\tau |) +\exp(-2\pi \Gamma _2| \tau |)\Big) \int_{0 }^{\infty } {\rm d} \lambda  b(\lambda )  \exp(-\pi \lambda (\Gamma _1+\Gamma _2) )\nonumber
\right \rangle 
    \\ \hspace{-0.3in}&&
 ~ +(2\pi )^4  \left \langle \Gamma _{1a}\Gamma _{1b}\Gamma _{2a}\Gamma _{2b} 
 \int_0^{\infty } \lambda {\rm d}\lambda  \label{Eq:G_4irrtimeAApp} 
b(\lambda + |\tau |) 
  \exp (-\pi(\Gamma _1+\Gamma _2) [\lambda + |\tau |])\right \rangle  \Big \}.~~
  \end{eqnarray}
The Fourier transform  of $\mathcal{H}^{abcd}_4(\epsilon )$ of Eq.  (\ref{Eq:H_4App}) is
    \begin{eqnarray}
     \hspace{-0.3in}&& 
      \tilde{\mathcal{H}}_4^{abcd} (\tau )= \label{Eq:H_4restBcollected}   -  (2\pi )^3\delta _{ab} \Big \langle  {\Gamma _{1a}\Gamma _{2b}\Gamma _{2c}\Gamma _{2d}\over \Gamma _2 }\times
      \\ \hspace{-0.3in}&&
      \nonumber \left[ \exp(-2\pi  \Gamma _2|\tau |  ) \int_0^{\infty } {\rm d}\lambda  b(\lambda)\exp (-\pi(\Gamma _1+\Gamma _2) \lambda ) 
 +\int_{|\tau |}^{\infty } {\rm d} \lambda b (\lambda  ) \exp(-\pi(\Gamma _1+\Gamma _2) \lambda ) \right] \Big \rangle   +(ab) \leftrightarrow  (cd). 
    \end{eqnarray}
The Fourier transform of the three-point function $\mathcal {F}_3^{abcd}$ in Eq.~(\ref{Eq:F_3App}) is
       \begin{eqnarray} 
        \hspace{-0.3in}&& 
      \tilde {\mathcal{F}}_3^{abcd}(\tau )=\label{Eq:F-3(fulltimeAApp}
      \\
  \hspace{-0.3in}&&
  -2 (1-\langle  S_{aa}\rangle) \delta _{ab} \tilde {C}_{acd}^{(3)}(\tau )\nonumber
       +     (ab) \leftrightarrow (cd) =  \nonumber
        \\
  \hspace{-0.3in}&& \nonumber
     + (2\pi )^2 (1-\langle S_{aa}\rangle) \delta _{ab} 
\Big\{\left \langle { \Gamma _{1a}\Gamma _{1c}\Gamma _{1d}\over \Gamma _1 }\exp (-2\pi \Gamma _1|\tau| )
\right \rangle   - \nonumber 
  b(|\tau |)
  \left \langle  { \Gamma _{1a}\Gamma _{2c}\Gamma _{2d}\over \Gamma _2  }\exp(-\pi(\Gamma _1+\Gamma _2) |\tau |)\right \rangle 
  \\        \hspace{-0.3in}&&  -    2\pi  \delta _{cd} 
      \left \langle  \Gamma _{1a}\Gamma _{1c}\Gamma _{2d}         \nonumber  \left[ \exp(-2\pi  \Gamma _1|\tau |  )  \int_0^{\infty } {\rm d}\lambda   b(\lambda) \exp (-\pi(\Gamma _1+\Gamma _2) \lambda) 
 + \int_{ |\tau|}^{\infty } {\rm d}\lambda  b( \lambda ) \exp(-\pi(\Gamma _1+\Gamma _2) \lambda ) \right] \right \rangle \Big\}   \nonumber 
\\  \hspace{-0.3in}&& +
(ab) \leftrightarrow (cd).     \nonumber 
      \end{eqnarray}
%$$$$$
\subsection {Qualitative Contributions to the Cross-Section Variance \label{App:qualitativevariance}}
%$$$$$
         Estimates concerning the relative importance of the different contributions in the SBW model are obtained by replacing the different total widths $\Gamma _1, \Gamma _2, \Gamma $ by a typical  total width $\Gamma _W$ in Eqs.~(\ref{Eq:F_4App})-(\ref{Eq:F_3App}) and Eqs.~(\ref{Eq:2pointfullenergyab}),~(\ref{Eq:2pointfullenergyaa}). Here, for simplicity we set $d=1$. The approximation assumes a negligible contribution from the partial widths with labels $a,b$ to the total width. The two-level form factor is given by Eq.~(\ref{Eq:clusterapprox}). The ratio of the variance to the square of the $S$-matrix autocorrelation function takes the following form for the inelastic case 
   \begin{eqnarray}
   \hspace{-0.3in}   
    C _{a\neq b}   (0 )/|C^{(2)}_{a\neq b}(0)|^2&=&
\label{Eq:inelvariancesimplifiedreduced}
 \Big \{  \mathcal {F}_4^{a\neq ba\neq b}(\epsilon =0) +\mathcal {G}_4^{a\neq ba\neq b}(\epsilon=0) + \Big | C ^{(0)}_{a\neq b} \Big |^2  %[S_{ab}^{*} S_{ab}^{}](\epsilon =
 (0 )|^2 \Big \} / \Big {|}C^{(2)}_{a\neq b}(0)\Big| | ^2 %C [S_{ab}^{*} S_{ab}^{}](\epsilon =0 )|^2
         \\    \hspace{-0.3in}
   &=& \left[\left( { 9 \over \pi \Gamma _W} - 
  {1\over (1+\pi  \Gamma _W)} 
 \right)-
\left( { 1\over (1+\pi  \Gamma _W )} +
  {( \pi  \Gamma _W )^2 \over (1+\pi  \Gamma _W )^3 }\right) 
 1 \right] \nonumber\\
&\longrightarrow&\left\{\begin{array}{ccc}
{ 9 \over \pi \Gamma _W} -1&{\rm for} &\pi\Gamma_W\lesssim 1\\
&&\\
{7 \over \pi \Gamma _W}+1&{\rm for} &\pi\Gamma_W\gtrsim 1\\
\end{array}\right.\nonumber
\end{eqnarray} 
The corresponding estimate for the elastic case equals
      \begin{eqnarray}
   \hspace{-0.3in}
    C_{aa} (0 )/  |C^{(2)}_{aa}(0)|^2 &=& \label{Eq:elvariancesimplifiedreduced}
    \Big\{ \mathcal{F}_4^{aaaa}(\epsilon =0)+\mathcal{G}_4^{aaaa}(\epsilon =0)+ 
  \mathcal{H}_4^{aaaa}(\epsilon =0 )   +
 Re~ ~\mathcal{F}_3^{aaaa} (\epsilon =0 )  
    \\ \hspace{-0.3in} &+& \nonumber
\left | C^{(2)}_{aa}(0)\right |^2+2(1-\langle S_{aa}\rangle) ^2 C^{(2)}_{aa}(0)\Big \} / \Big|C^{(2)}_{aa}(0)\Big|^2 
\\   \hspace{-0.3in}&\simeq & %%%%%%%%%%%%%%%%%%%%%%%%%%%%%%%%%%%%%%%%       F4 term
\Big \{\Big[ {105 \over \pi\Gamma _W}   - \label{Eq:approximateelasticselfcorrelationF_4}
   { 9 \over (1+\pi \Gamma _W )}    \Big] 
-9  \Big [      {1 \over (1+\pi \Gamma _W) }      ~ +{(\pi \Gamma _W)^2  \  \over (1+\pi \Gamma _W)^3 } \Big ]-
% %%%%%%%%%%%%%%%%%%%%%%%%%%%%%%%%%%%%%%%%%%%%%       H4 term
       30   %  factor 2 is corrected replacing 60 by 30 
        \left[ {1\over  (1+\pi \Gamma _W) }+{\pi \Gamma _W \over (1+\pi \Gamma _W) ^2}  \right] \nonumber
   \\ \hspace{-0.3in} &+&  %%%%%%%%%%%%%%%%%%%%%%%%%%%%%%%%%%%%%%%%%%%%%%  F3  term
     \left({ 2\over \nonumber
      <  S_{aa} > +1}\right) 2\label{Eq:F-3(fulltimeAApp1}   										   
\Big\{  15      - 
     { 3\pi\Gamma _W  \over (1+\pi \Gamma _W)    }  -       3        \left[  { \pi\Gamma _W\over   
          (1+\pi \Gamma _W)}    +{(\pi\Gamma _W)^2 \over   (1+\pi \Gamma _W)^2}   \right]    \Big\}  
     \\        \hspace{-0.3in} &+&  \nonumber			
      \left( 3 - { \pi\Gamma _W \over (1+\pi \Gamma _W )}      \right)^2 +
  %%%%%%%%%%%%%%%%%%%%%%%%%%%%%%%%%%%%%%%%%%%%%%%%%%%%%%%%%%%%%% %	C[S_{a a}S_{aa}^*](\epsilon=0) term
   \left({2\over <S_{aa}> +1}\right)^{2}  4  \pi
  \Gamma _W	\Big[  3    -   
    { \pi\Gamma _W\over  (1+\pi \Gamma _W )}    \Big] \Big \}   \Big /  \left( 3 - { \pi\Gamma _W \over (1+\pi \Gamma _W )}  
    \right)^2 
    \nonumber \\
&\longrightarrow&\left\{\begin{array}{ccc}
{35\over 3\pi\Gamma _W}-1&{\rm for} &\pi\Gamma_W\lesssim 1\\
&&\\
4+2\pi\Gamma _W&{\rm for} &\pi\Gamma_W\gtrsim 1
\end{array}\right.\nonumber
\end{eqnarray}
Note, that the typical scale of the Weisskopf unit in this context is $\pi\Gamma_W$ and not $\Gamma_W$.
   \end{widetext}

 %AAAAAAAAAAAAAAAAAAAAAAAAAAAAAAAAAAAA


\begin{thebibliography}{99}
\bibitem{footnote}
Equation (12) represents the simplest version of SBW type pole models. Unitarity consistency on the average is obtained by adding an appropriate constant to each diagonal pole term such that each $\langle S_{aa}\rangle$ has the value consistent with its transmission coefficient Eq.~(3). This modification is imposed in Sects. III and following. It has no effect on the so-called irreducible parts of the correlations.
\end{thebibliography}
\end{document}